\documentclass{article}
\usepackage{arxiv}

\usepackage[utf8]{inputenc} 
\usepackage[T1]{fontenc}    
\usepackage{commath}
\usepackage{hyperref}       
\usepackage{url}            
\usepackage{mathtools}
\usepackage{booktabs}       
\usepackage{amsfonts}       
\usepackage{nicefrac}       
\usepackage{microtype}      
\usepackage{lipsum}
\usepackage{graphicx}
\graphicspath{ {./images/} }
\usepackage{color}
\usepackage{tikz}
\usepackage{esint} 
\usepackage{caption}
\usepackage{subcaption}
\usepackage{graphicx}
\usepackage{multirow}
\usepackage{array}
\usepackage[flushleft]{threeparttable}
\usepackage{adjustbox}
\usepackage[nointegrals]{wasysym}
\usepackage{float}

\newtheorem{theorem}{Theorem}

\newtheorem{ack}{ACKNOWLEDGEMENTS}
\DeclareUnicodeCharacter{200B}{I~AM~HERE!!!!}

\usetikzlibrary{shapes,decorations,arrows,calc,arrows.meta,fit,positioning}
\tikzset{
    -Latex,auto,node distance =1 cm and 1 cm,semithick,
    state/.style ={ellipse, draw, minimum width = 0.7 cm},
    point/.style = {circle, draw, inner sep=0.04cm,fill,node contents={}},
    bidirected/.style={Latex-Latex,dashed},
    el/.style = {inner sep=2pt, align=left, sloped}
}

\usepackage{amssymb}
\usepackage{pifont}

\usepackage{latexsym}
\usepackage{savesym}
\usepackage{mathtools}
\savesymbol{iint}
\usepackage{txfonts}
\restoresymbol{TXF}{iint}
\usepackage{amsfonts}
\usepackage{amssymb}
\usepackage{psfrag}
\usepackage{graphicx}
\usepackage{subcaption}
\usepackage{mathrsfs} 
\usepackage{caption}
\usepackage{subcaption}

\title{ Doubly-Robust Dynamic Treatment Regimen Estimation with Binary Outcomes}

\author{
Cong Jiang, Michael Wallace, Mary Thompson \\
  Department of Statistics and Actuarial Science\\
  University of Waterloo\\ \texttt{\{cong.jiang, michael.wallace, methompson\} @uwaterloo.ca}\\
}
\date{}

\begin{document}
\maketitle
\begin{abstract}
In precision medicine, Dynamic Treatment Regimes (DTRs) are treatment protocols that adapt over time in response to a patient's observed characteristics. A DTR is a set of decision functions that takes an individual patient's information as arguments and outputs an action to be taken. Building on observed data, the aim is to identify the DTR that optimizes expected patient outcomes. Multiple methods have been proposed for optimal DTR estimation with continuous outcomes. However, optimal DTR estimation with binary outcomes is more complicated and has received comparatively little attention. Solving a system of weighted generalized estimating equations, we propose a new balancing weight criterion to overcome the misspecification of generalized linear models’ nuisance components. We construct binary pseudo-outcomes, and develop a doubly-robust and easy-to-use method to estimate an optimal DTR with binary outcomes. We also outline the underlying theory, which relies on the balancing property of the weights; provide simulation studies that verify the double-robustness of our method; and illustrate the method in studying the effects of e-cigarette usage on smoking cessation, using  observational data from the Population Assessment of Tobacco and Health (PATH) study. 
\end{abstract}

\section{Introduction}
\emph{Precision medicine} (also known as personalized medicine) refers to the principle of tailoring treatment according to patients’ individual characteristics. Different from traditional pharmacological practice, where treatments are recommended based solely on the disease diagnosis, precision medicine uses patient information to make a patient-specific treatment recommendation. \emph{Dynamic treatment regimes (DTRs)}, as a formal statistical analysis framework for precision medicine, are sequential decision rules that take patients’ individual information as input, and output individualized treatment recommendations. Identifying the \emph{optimal} DTRs, which are sequences of treatment rules that yield the best-expected health outcome across a population, is a challenging but key task in the process of personalizing treatment.

Most optimal DTR estimation methodologies focus on continuous outcomes. These include regression-based methods such as Q-learning (\cite{sutton2018reinforcement}), G-estimation (\cite{robins2004optimal}) and dynamic weighted ordinary least squares (dWOLS, \cite{wallace2015doubly}), and value-search methods such as (augmented) inverse probability of treatment weighting (\cite{zhang2013robust}) and outcome weighted learning (\cite{zhao2012estimating}). As a continuous-outcome problem counterpart, the discrete-outcome problem is more challenging, yet valuable in real-world applications. Binary outcomes, such as treatment failure or success, are crucial measures in many medical or health studies. However, to date, optimal DTR estimation for binary outcomes has received little attention. 

Although most methodologies target continuous outcomes, there are some existing studies regarding non-continuous outcomes. Some theoretical developments in the DTR literature have focused on discrete-outcome settings, including Q-learning with discrete outcomes (Bernoulli and Poisson) utilities (\cite{moodie2014q}), an extension of G-estimation to the case of non-additive treatment effects for discrete outcomes (\cite{wallace2019model}), and the extension of dWOLS to time-to-event data with survival outcomes subject to right-censoring (\cite{simoneau2020estimating}). 

For binary outcomes, the recently proposed DTR estimation approaches are reliant on either Q-learning, which offers relatively straightforward implementation, or G-estimation, which is \emph{doubly robust} in the sense of offering a consistent estimator of a treatment effect if at least one of two nuisance models is correctly specified. For example, considering cases of cancer and graft-versus-host disease treatment, to maximize the probability of the binary outcome of two-year disease-free survival, Moodie and Krakow \cite{moodie2020precision} implemented Q-learning in a multi-stage treatment decision analysis, employing logistic regression at each stage. This method was shown to be easy to implement, but suffered from problems of sensitivity to misspecification of the outcome model. Wang et al. \cite{wang2017congenial}, meanwhile, proposed a G-estimation based method for binary multiplicative structural nested mean models. They constructed a locally semiparametric efficient estimator, and demonstrated that such estimators boast the aforementioned double robustness property. Recently, analyzing micro-randomized trial data with binary outcomes, Qian et al. \cite{qian2019estimating} defined the causal excursion effect, which refers to a log relative risk between two excursions from a treatment protocol, and also provided a semiparametric and locally efficient estimator of the causal excursion effect. Based on the research of Robins (\cite{robins1994estimation}, \cite{robins2004optimal}), these two semiparametric methods are doubly robust. However, their drawbacks include complexity of theory and implementation which may prove challenging for many practitioners. 

In contrast to Q-learning and G-estimation, dynamic weighted ordinary least squares offers an approach to DTR estimation that is doubly robust while also relatively straigthforward to implement. In the case of identifying a multi-stage DTR (that is, a sequence of treatment decisions at fixed time points), dWOLS proceeds via a sequence of weighted ordinary least squares regressions. Building on this methodology, we propose the dynamic weighted generalized linear model (dWGLM); an extension of dWOLS to the case of binary outcomes that offers similar properties in terms of robustness and ease of implementation.


This paper is organized as follows: Section 2 introduces the proposed doubly robust regression-based DTR estimation framework with binary outcomes, where we take the term doubly robust to include the approximate consistency of the estimator; Section 3 describes simulation studies, demonstrating the double robustness of our methods; Section 4 illustrates our methodology using observational data from the Population Assessment of Tobacco and Health (PATH) study, and Section 5 concludes with a discussion.


\section{Methodology}
\subsection{Introductory notation and settings}

We first introduce the necessary notation and settings in the standard DTR literature. Then, we provide a summary of our specific estimation procedures. Note that we suppress patient-level notation for ease of exposition. Typically, we assume that a DTR contains a total of $K$ treatment stages in a multiple-stage treatment decision problem.  Let $Y$ denote the patient outcome, a binary variable that takes values in $\{0, 1\}$; it is observed after the assignment of all the treatments. We also assume that $Y = 1$ is preferred. Let $a_j$ be the value of the $j^{th}$ stage treatment decision, or action, $A_j$ for $j = 1,..., K$; it is also a binary variable with 0 referring to a baseeline treatment, such as standard care or no treatment. Let $\boldsymbol{x}_j$ denote the observed value of the covariate vector $\boldsymbol{X}_j$ of non-treatment information, such as age, disease severity, response to previous treatments, and so on; it is recorded at Stage $j$, and is known preceding the $j^{th}$ stage treatment decision. Finally, let $\boldsymbol{h}_j$ be the value of the covariate matrix $\boldsymbol{H}_j$; it includes all patient information history preceding the $j^{th}$ stage treatment decision, and  can contain non-treatment information up to Stage $j$ (i.e., $\boldsymbol{x}_1, \boldsymbol{x}_2, ... , \boldsymbol{x}_j$) along with previous treatments ($a_1, ... , a_{j-1}$). 

In addition, over- and underline are used to represent the past and future, respectively. For instance, $\bar{a}_{j}$ indicates a vector of the first $j$ treatment decisions (i.e., $a_1, a_2, . . ., a_j$) and $\underline{a}_{j+1}$ denotes the vector of treatment decisions from Stage $j + 1$ onwards (i.e., $a_{j+1}, a_{j+2}, . . ., a_K$). Therefore, the history prior to the $j^{th}$ treatment decision can be written as $\boldsymbol{h}_{j}=\left(\overline{\boldsymbol{x}}_{j}, \overline{a}_{j-1}\right)$. The Stage $j$ optimal treatment is denoted as $a_j^{opt}$. 

A fundamental component of DTR estimation is the \emph{blip function} (\cite{robins2004optimal}). Denoting $Y^*(\boldsymbol{a})$ as the potential (or counterfactual) outcome under treatment regime $\boldsymbol{a}$, we then define the optimal \emph{blip-to-reference function} for Stage $j$ as: 
\begin{equation}\label{blip}
    \gamma_{j}\left(\boldsymbol{h}_{j}, a_{j}\right) = g\left(\mathbb{P}\left[Y^{*}\left(\bar{a}_{j}, \underline{a}_{j+1}^{o p t}\right) = 1 \mid \boldsymbol{H}_{j}=\boldsymbol{h}_{j}\right]\right) - g\left(\mathbb{P}\left[Y^{*}\left(\bar{a}_{j-1}, a_{j}^{ref}, \underline{a}_{j+1}^{o p t}\right) =1 \mid \boldsymbol{H}_{j}=\boldsymbol{h}_{j}\right]\right),
\end{equation}
which is the difference in the $g$ link function transformation of the mean of the binary outcome when using a reference treatment $a_j^{ref}$ (such as a control) instead of $a_j$ at Stage $j$, in individuals with history $\boldsymbol{h}_j$ who are subsequently optimally treated (i.e., receiving $\underline{a}_{j+1}^{o p t}$ ). Note that, for binary outcomes, there are several options for the \emph{link function} $g: (0, 1) \rightarrow (-\infty, +\infty)$, and if $g$ is the canonical (logit) link for the binomial, i.e., $g(p)=log(p/(1-p))$, the equation (\ref{blip}) blip function then denotes the log odds ratio of expected potential binary outcomes for patients who are treated versus untreated. When $g$ is the identity link $g(p) = p$, the blip function represents the risk difference of expected potential binary outcomes for patients who are treated versus untreated. Other choices of link function could be the probit link $g(p) = \Phi^{-1}(p)$, where $\Phi$ is the cumulative distribution function (C.D.F.) of the standard normal distribution, and the complementary log-log function $g(p) = log(-log(1-p))$. Further, the \emph{robit link}, $g(p) = F_{\nu}^{-1}(p)$ where $F_{\nu}(x)$ is the C.D.F. of the Student's $\bold{t}$-distribution with center zero, scale parameter one, and $\nu$ degrees of freedom, could be considered. Some appealing properties of the robit link have been identified. For example, the robit link can be considered as a generalization of probit link and an approximate generalization of logit link, and it provides a robust estimation in that the coefficient estimates are less influenced by individual outlying data points (\cite{liu2004robit}).

The outcome probabilities can be framed in the context of blip functions such that $g[\mathbb{P}(Y = 1)] = g[\mathbb{P}(Y^{opt} = 1)] - \sum_{j=1}^{K}\left[\gamma_{j}\left(\boldsymbol{h}_{j}, a_{j}^{o p t}\right)-\gamma_{j}\left(\boldsymbol{h}_{j}, a_{j}\right)\right]$. A corresponding concept to blips is \emph{regrets} (\cite{murphy2003optimal}). The regret function (for Stage $j$) can be defined as $$\mu_{j}\left(\boldsymbol{h}_{j}, a_{j}\right)=g\left(\mathbb{P}\left[Y^{*}\left(\bar{a}_{j-1}, \underline{a}_{j}^{o p t}\right) = 1 \mid \boldsymbol{H}_{j}=\boldsymbol{h}_{j}\right]\right) - g\left(\mathbb{P}\left[Y^{*}\left(\bar{a}_{j}, \underline{a}_{j+1}^{o p t}\right) = 1 \mid \boldsymbol{H}_{j}=\boldsymbol{h}_{j}\right]\right),$$ which is the $g$ (link) function transformation of expected loss or regret arising from prescribing treatment $a_j$ at Stage $j$ instead of the optimal treatment  $a^
{opt}_j$, assuming optimal treatment is received in the later stages. Thus, the blip and regret functions can be related such that $\mu_{j}\left(\boldsymbol{h}_{j}, a_{j}\right)=\gamma_{j}\left(\boldsymbol{h}_{j}, a_{j}^{o p t}\right)-\gamma_{j}\left(\boldsymbol{h}_{j}, a_{j}\right)$, showing that, if the individual was optimally treated, the regret function will equal 0. Otherwise, the regret function will be positive. The positive value of regrets represents the expected improvement in outcome had the patient received optimal treatment compared to the observed treatment. The concept of regrets provide a distinct view to consider the effects of the treatment in terms of the optimal treatment, and it also simplifies some expressions in later sections. 

Building on the concept of the blip and regret functions, the main goal of DTR estimation is to identify the optimal treatment decision that maximizes the blip function or equivalently minimizes the regret function. For instance,  we consider the outcome model that can be decomposed into two components: $g(\mathbb{E}\left[Y^{*}(\boldsymbol{a}) | \boldsymbol{H}=\boldsymbol{h}\right])=\sum_{j=1}^{K}\left[f_j\left(\boldsymbol{h}^{\beta}_{j}; \boldsymbol{\beta}\right)+ \gamma_{j}(\boldsymbol{h}_{j}^{\psi}, a_{j}; \boldsymbol{\psi}_{j}) \right],$ where 
$f_j\left(\boldsymbol{h}^{\beta}_{j}; \boldsymbol{\beta}\right)$ and $\gamma_{j}(\boldsymbol{h}_{j}^{\psi}, a_{j}; \boldsymbol{\psi}_{j})$ are so-called \emph{treatment-free} and
blip models, respectively, and $\boldsymbol{h}_{j}^{\beta}$ and $\boldsymbol{h}_{j}^{\psi}$ are subsets of covariates in $\boldsymbol{h}_{j}$. Note that $f_j\left(\boldsymbol{h}^{\beta}_{j}; \boldsymbol{\beta}\right)$ denotes the expected contribution of covariates $\boldsymbol{h}_{j}^{\beta}$ (often termed \emph{predictive variables}) in the absence of treatment for Stage $j$. The treatment-free model is unrelated to making decisions about optimal treatment selections; therefore, the optimal treatment is that which maximizes the blip (or minimizes the regret). By the definition of the blip function, at each treatment decision point, $\gamma_{j}(\boldsymbol{h}_{j}^{\psi}, 0 ; \boldsymbol{\psi}_{j}) = 0$. Given the so-called \emph{prescriptive} or \emph{tailoring} variable $\boldsymbol{h}_{j}^{\psi}$, typically a small subset of $\boldsymbol{h}_{j}$, the optimal treatment decision for Stage $j$ is ``$a^{opt}_j = 1,\ if\  \gamma_{j}(\boldsymbol{h}_{j}^{\psi}, 1 ; \boldsymbol{\psi}_{j}) > 0; \ a^{opt}_j = 0,\ otherwise.$" Thus, to make the optimal treatment decision, it is sufficient to estimate the parameters in the blip model, and consider the parameters in the treatment-free model as nuisance parameters.

Finally, to proceed with DTR estimation for either observational or randomized data, we make the following identifiability assumptions: (1) consistency (\cite{rubin1980randomization}): the potential outcome under some sequence of treatments is equal to the observed outcome if those treatments were those actually received; (2) there are no unmeasured confounders for any possible treatment regimes (or sequential randomization assumption \cite{robins1986new}), i.e., conditional on current patient history, the current stage treatment is independent of future potential outcome or covariates; (3) no interference  between individuals (\cite{cox1958planning}): the outcome of one patient is unaffected by the treatment assignment of other patients, and (4) positivity (\cite{robins2004optimal}): at each decision point, there is a non-zero probability of being assigned to each of the treatment levels, no matter what the past treatment and covariate history.

\subsection{Q-learning with binary outcomes}
We first introduce Q-learning for binary outcomes as motivation and elucidation, then provide our proposed method. To identify the optimal DTRs, Q-learning recursively solves treatment decision problems starting from the last stage, and at each stage, the \emph{Q-function} is defined as follows (\cite{moodie2014q},\cite{moodie2020precision}):
$$Q_{K}\left(\boldsymbol{h}_{K}, a_{K}\right)= g\left(\mathbb{P}\left[Y^*(a_{K}) = 1 \mid \boldsymbol{H}_{K}=\boldsymbol{h}_{K}, A_{K} = a_{K}\right]\right);$$ and
$$\begin{aligned} Q_{j}\left(\boldsymbol{h}_{j}, a_{j}\right)= g\left( \mathbb{E}\left[\max _{A_{j+1}} g^{-1}\left[Q_{j+1}\left(\boldsymbol{H}_{j+1}, A_{j+1}\right)\right]  \mid \boldsymbol{H}_{j}=\boldsymbol{h}_{j}, A_{j}=a_{j}\right]\right)  & \text { for } j<K. \end{aligned}$$

Suppose the Q-functions are modeled linearly such that $Q_{j}\left(\boldsymbol{h}_{j}, a_{j} ; \boldsymbol{\beta}_{j}, \boldsymbol{\psi}_{j}\right)=\boldsymbol{\beta}_{j}^{\top} \boldsymbol{h}_{j}^{\beta}+\boldsymbol{\psi}_{j}^{\top} a_{j} \boldsymbol{h}_{j}^{\psi},$ where $\boldsymbol{h}_{j}^{\beta}$ and $\boldsymbol{h}_{j}^{\psi}$ are subsets of covariates in $\boldsymbol{h}_{j}$. Note that the pseudo-outcome-probability, because of the monotone increasing property of the $g^{-1}$ function, $$\max_{A_{j+1}}g^{-1}\left[ Q_{j+1}\left(\boldsymbol{H}_{j+1}, A_{j+1}\right) \right]= g^{-1}\left[\max_{A_{j+1}} Q_{j+1}\left(\boldsymbol{H}_{j+1}, A_{j+1}\right)\right],$$ refers to the "best possible" probability of the outcome a patient could have based on the proposed outcome models in the preceding stage.  Then the treatment decisions are made according to the estimates of $\boldsymbol{\psi}$ in each stage.  For example, in Stage $j$, "$ If \ \boldsymbol{\hat{\psi}}_{j}^{\top} \boldsymbol{h}_{j}^{\psi}>0,\ treat; otherwise,\ leave\ untreated.$" However, we note that we should correctly specify all Q-function models, including the treatment-free models, to acquire consistent estimators of $\boldsymbol{\psi}$.

\subsection{Balancing property}
In the previous sub-section, Q-learning provides a comparatively simple to follow method via a generalized linear model with binary outcomes, such as logistic regression, but it lacks robustness to misspecification of treatment-free models. Our proposed approach, inspired by dWOLS, employs balancing weights to overcome the possible misspecification of these models.

dWOLS employs a series of sequential weighted regressions to consistently estimate the parameters of interest in the outcome model. The double robustness of dWOLS relies on balancing weights, which are a function of the propensity score and thus determined by the underlying treatment model.  We define the propensity score (\cite{rosenbaum1983central}) as $\pi(\boldsymbol{x}) :=\mathbb{P}(A = 1 \mid \boldsymbol{x})$. Then the balancing weights criterion introduced in dWOLS establish independence between the covariates and treatment in the weighted dataset. Thus, the bias in estimating the blip parameter, introduced due to the dependence between covariates and treatment is removed.  We denote by $w^{d}$ a choice of dWOLS balancing weights that satisfy the criterion $(1 - \pi(\boldsymbol{x}))w(0, \boldsymbol{x}) =  \pi(\boldsymbol{x}) w(1, \boldsymbol{x})$ as proposed in Theorem 1 of \cite{wallace2015doubly}.  For efficient estimation, Wallace and Moodie \cite{wallace2015doubly} suggested the use of  "absolute value" weights of the form $w = |A - \mathbb{E}(A \mid \boldsymbol{x}) |$, also called "overlap weights", which have been extensively discussed by Li et al. \cite{li2018balancing}.

Suppose that $\boldsymbol{x}^{\beta}$ and $\boldsymbol{x}^{\psi}$ are two subsets of the covariates included in $\boldsymbol{x}$, and that the true outcome model is $ g\left(\mathbb{P} [ Y = 1 | \boldsymbol{x} , a]\right) = f (\boldsymbol{x}_{}^{\beta}; \boldsymbol{\beta}) + \gamma(\boldsymbol{x}^{\psi}, a; \boldsymbol{\psi}),$ where $\gamma(\boldsymbol{x}^{\psi}, a; \boldsymbol{\psi})$ is in the linear form $\boldsymbol{\psi}^{\top}a \boldsymbol{x}^{\psi}$ but $f$ is an arbitrary function, and $g$ is the link function for binary outcomes. Then, the following theorem holds.

\begin{theorem}{Balancing property for GLM with binary outcomes}\label{P2thm1}

When the true outcome model satisfies $ g\left(\mathbb{P} [ Y = 1 | \boldsymbol{x} , a]\right) = f (\boldsymbol{x}_{}^{\beta}; \boldsymbol{\beta}) + \gamma(\boldsymbol{x}^{\psi}, a; \boldsymbol{\psi}),$ where $g$ is the link function, a weighted generalized linear model based on the corresponding linear predictor will yield approximately consistent estimators of $\boldsymbol{\psi}$ if the weights satisfy 
\begin{equation} \label{wtcri}
 (1- \pi(\boldsymbol{x})) w(0, \boldsymbol{x})\kappa(0, \boldsymbol{x}) = \pi(\boldsymbol{x}) w(1, \boldsymbol{x})\kappa(1, \boldsymbol{x})
\end{equation}
 where $\kappa(a, \boldsymbol{x}) = {g^{-1}}^{\prime}(\boldsymbol {{\beta}^{*} }^{\top} \boldsymbol{x}^{\beta} + \boldsymbol { \psi^* }^{\top} a\boldsymbol{x}^{\psi})$ and ${g^{-1}}^{\prime} (x) $ is the first derivative of the  inverse link function (i.e., ${{g^{-1}}^{\prime}}(x) = \frac{d g^{-1}(x)}{dx}$), and $\boldsymbol { \beta^* }$ and  $\boldsymbol { \psi^* }$ are defined through $\mathbb{E}\left((1 - \pi(\boldsymbol{x}))\boldsymbol{x}w^{d}(0, \boldsymbol{x})\left[g^{-1}(f(\boldsymbol{x}^{\beta})) - g^{-1}(  \boldsymbol { \beta^* }^{\top} \boldsymbol{x}^{\beta})\right] \right) = \boldsymbol{0}$ and $ \mathbb{E}\left( \pi(\boldsymbol{x})\boldsymbol{x} w^{d}(1, \boldsymbol{x})\left[g^{-1}(f(\boldsymbol{x}^{\beta}) + \boldsymbol { \psi^* }^{\top} \boldsymbol{x}^{}) - g^{-1}(  \boldsymbol { \beta^* }^{\top} \boldsymbol{x}^{\beta}+ \boldsymbol { \psi^* }^{\top} \boldsymbol{x}^{})\right]\right) = \boldsymbol{0}$, respectively.
 \end{theorem}
\textbf{Proof of Theorem 1}: See Appendix \ref{Appx.P2A1}. In addition, building on the systems of estimating equations in this proof, Appendix \ref{Appx.P2B1} presents the proof of the uniqueness of $\boldsymbol{\beta^*}$. 

The balancing weights criterion (Equation \ref{wtcri}) is similar to that used in dWOLS, both being built on propensity scores. However, the Equation \ref{wtcri} contains an extra term $\kappa(a, \boldsymbol{x})$. This $\kappa(a, \boldsymbol{x})$ is related to the $g$ link function, and it is derived, as shown in the proof \ref{Appx.P2A1}, from the system of estimation functions of the GLMs for the purpose of addressing the misspecification of the treatment-free model. It is important to emphasize that we only employ a linear predictor for any form of the true treatment-free model in a GLM. Even if the true treatment-free model is non-linear, the estimator of $\boldsymbol{\psi}$ is guaranteed to be "approximately" consistent,  where the use of "approximately" originates from the error term in the Taylor series approximation of the inverse link function (see proof of Theorem 1 in Appendix \ref{Appx.P2A1}).  That is, by the Taylor expansion of the inverse link function ($g^{-1}$), we can connect the two estimation equations of two different treatment groups, and thus construct the corresponding balancing weights criterion.  Accordingly, the Taylor expansion induces an error term, and this error term will be small when a linear predictor tends to vary in an interval where $g^{-1}$ is approximately linear. In some applications, to acquire a more-accurate estimation, it may be possible to choose the range of covariates so that the linear predictor varies in such an interval.

Therefore, from the standpoint of robust estimation of the GLM with $g$ link, we call this $\kappa(a, \boldsymbol{x})$ an "adjustment" factor in the balancing weights criterion for binary outcomes. The "adjustment" factor is a function of the linear predictor $\boldsymbol {{\beta}^{*} }^{\top} \boldsymbol{x}^{\beta} + \boldsymbol { \psi^* }^{\top} a\boldsymbol{x}^{\psi}$, where $\boldsymbol { \beta^* }$ and  $\boldsymbol { \psi^* }$ are roots of the estimating functions of the GLMs with standard dWOLS weights $w^{d}$. Consequently, to construct the balancing weights for GLM with binary outcomes, two crucial steps are required: (1) identify the "adjustment" factor by conducting a weighted GLM (e.g., logistic regression) with the standard dWOLS weights; (2) compute the balancing weights based on the estimated "adjustment" factor, propensity score, and weights criterion \ref{wtcri}.  These two steps are illustrated in our proposed method of optimal DTR estimation with binary outcomes in the next section.

\subsection{Dynamic weighted generalized linear model}
Inspired by the easy implementation of Q-learning and the double robustness property of G-estimation and dWOLS, our proposed method, the dynamic weighted generalized linear model (dWGLM), estimates the blip parameters in terms of binary treatments and outcomes. Note that, for the multiple-stage decision problems, because of prognostic effects and delayed treatment effects (\cite{kosorok2019precision}), the current treatment decisions will not only affect the intermediate outcomes but also affect the future ones; thus, the decisions should be "farsighted". Backward induction is used in Q-learning for sequential decision problems. Thus the sequential decision problems can be divided into a set of single-stage problems, each of which aims to optimize the stage specific \emph{pseudo-outcomes}: the potential outcomes if the patients were treated - possibly contrary to fact - optimally at subsequent stages. Similar to the process of Q-learning with binary outcomes, dWGLM involves a series of weighted generalized linear models of either the observed outcome $y$ (at Stage $K$) or \emph{binary pseudo-outcomes} $\widetilde{\mathcal{Y}_j}$ (for stages $j < K$) on subject histories. These binary pseudo-outcomes are random variables from the Bernoulli distribution with success probability $\mathbb{P}(\widetilde{\mathcal{Y}_j}=1) = g^{-1}\left[g[\mathbb{P}(Y = 1\mid \boldsymbol{h}_{K}, a_{K}; \boldsymbol{\hat{\beta}}_{K}, \boldsymbol{\hat{\psi}}_{K})] + \sum_{k=j+1}^{K} \mu_{k}\left(\boldsymbol{h}_{k}, a_{k} ; \boldsymbol{\hat{\psi}}_{k}\right)\right].$ In continuous outcome G-estimation and dWOLS settings, the pseudo-outcome definition relies on the final observed outcome $y$, that is, $\widetilde{\mathcal{Y}_j} = y + \sum_{k=j+1}^{K} \mu_{k}\left(\boldsymbol{h}_{k}, a_{k} ; \boldsymbol{\hat{\psi}}_{k}\right).$ For the binary case, however, we concentrate on probabilities of the outcome being one rather than directly using the observed binary outcome $y$. Therefore, we employ the last stage model to estimate $\mathbb{P}(Y = 1 \mid \boldsymbol{h}_{K}, a_{K})$, and then combine the regrets to acquire the pseudo-outcome probabilities.  We also emphasize that, in keeping with the goal of GLM, we mainly focus on modeling the probability that the pseudo-outcome equals one, i.e., $\mathbb{P}(\widetilde{\mathcal{Y}_{j}}=1 \mid \boldsymbol{h}_{j}, a_{j}),$ for each stage ($j = 1,2,...,K$). Moreover, to improve the efficiency of $\boldsymbol{\psi}$ estimators, we construct the $\widetilde{\mathcal{Y}}$ multiple times (say $R$ times) in each stage, and implement the estimation $R$ times in each stage. 
Therefore, for the multistage decision analysis, the dWGLM procedure could be implemented by the following steps at each stage of the analysis, starting from the last stage $K$ and working backwards towards the first stage:
\begin{itemize}
    \item \textbf{Step 1:} Construct the stage $j$ \emph{pseudo-outcome}: set $\widetilde{\mathcal{Y}_j} =y $ if $j = K$. Otherwise, use prior estimates $\boldsymbol{\hat{\beta}}_{K}$ and $\boldsymbol{\hat{\underline{\psi}}}_{j +1} = (\boldsymbol{\hat{\psi}}_{j+1},..., \boldsymbol{\hat{\psi}}_{K})$ to randomly generate $\widetilde{\mathcal{Y}_j}$, which takes the value $1$ with probability $\mathbb{P}(\widetilde{\mathcal{Y}_j}=1) = g^{-1}\left[g[\mathbb{P}(Y= 1 \mid \boldsymbol{h}_{K}, a_{K}; \boldsymbol{\hat{\beta}}_{K}, \boldsymbol{\hat{\psi}}_{K})]  +\sum_{k=j+1}^{K} \mu_k (\boldsymbol{h}_{k}, a_{k} ; \boldsymbol{\hat{\psi}}_{k}) \right]$, $R$ times, to yield $\widetilde{\mathcal{Y}_j^{1}}, \widetilde{\mathcal{Y}_j^{2}}, ..., \widetilde{\mathcal{Y}_j^{R}}$.
    \item \textbf{Step 2:} Specify the stage $j$ treatment model $\mathbb{E}\left[A_{j} | \boldsymbol{h}_{j}^{\alpha}; \boldsymbol{\alpha}_j \right]$. The treatment model parameters $\boldsymbol{\alpha}_j$ (estimated, for example, via logistic regression) are used to compute a weight $w_j$, such as $w_{j}=\left|a_{j}-\mathbb{E}\left[A_{j} | \boldsymbol{h}_{j}^{\alpha}; \boldsymbol{\hat{\alpha}}_j \right]\right|$. 
    
    \item \textbf{Step 3:} Specify the stage $j$ treatment-free and blip models, and perform a weighted generalized linear model of $\widetilde{\mathcal{Y}_j^{r}}$ on the terms in the treatment-free and blip models, using weights from Step 2 to get estimates $\boldsymbol{\hat{\beta}}_{j}^{old, r}$, $\boldsymbol{\hat{\psi}}_{j}^{old, r}$ for $r = 1,..., R$; that is, for each $r = 1,..., R$, use the model \begin{align} \label{model3}
        g(\mathbb{E}[\widetilde{\mathcal{Y}_j^r} \mid a_j, \boldsymbol{h}_{j}; \boldsymbol{\beta}_j,  \boldsymbol{\psi}_j ]) = \boldsymbol{\beta}_{j}^{\top} \boldsymbol{h}_{j}^{\beta} + \boldsymbol{\psi}_{j}^{\top} a_{j} \boldsymbol{h}_{j}^{\psi}.
    \end{align}
    
    \item \textbf{Step 4:} Use $\boldsymbol{\hat{\beta}}_{j}^{old, r}$, $\boldsymbol{\hat{\psi}}_{j}^{old, r}$ from Step 3 to compute $$\kappa^{r}(a_j,\boldsymbol{h}_{j}) = {g^{-1}}^{\prime}(\boldsymbol{\hat{\beta}}_j^{{old, r}^{\top}} \boldsymbol{h}_{j}^{\beta} + \boldsymbol{\hat{\psi}}_j^{{old, r}^{\top}} a_j\boldsymbol{h}_{j}^{\psi}),$$ where ${g}^{-1}$ is identified based on the link function in Step 3. Then, construct the new weights 
    \begin{equation}\label{neww}
        w^{new, r}_j(a_j; \boldsymbol{h}_{j}) = |a_j - \mathop{\mathbb{E}[A_j | \boldsymbol{h}_{j}^{\alpha}]}| * \kappa^{r}(1 - a_j , \boldsymbol{h}_{j}^{}).
    \end{equation}
    \item \textbf{Step 5:} Perform a weighted GLM with the new weights (i.e., $w^{new, r}_j(a_j; \boldsymbol{h}_{j})$) to get revised estimates $\boldsymbol{\hat{\beta}}_{j}^{r}$, $\boldsymbol{\hat{\psi}}_{j}^{r}$ for each $r$. Estimate $\boldsymbol{\psi}_j$ by $\boldsymbol{\hat{\psi}}_{j} = R^{-1}\sum_{r} \boldsymbol{\hat{\psi}}_{j}^{r}$, then use parameter estimators  $\boldsymbol{\hat{\psi}}_{j}$ to construct the $j^{th}$ stage optimal treatment rule, which is $ prescribe\ a_j = 1\ if\  \boldsymbol{\hat{\psi}}_j^{\top}\boldsymbol{H}^{\psi}_j > 0; \ then\ prescribe\ a_j= 0\ otherwise$.  
    \item \textbf{Step 6:} Return to Step 1 and analyze Stage $j - 1$ if there are more stages to analyze.
\end{itemize}

Our proposed dWGLM approach thus contains at each stage a two-step GLM estimation process for binary outcomes. Each step uses GLM for binary outcomes (e.g., logistic regression) to estimate the parameters of interest. The first step could employ logistic regression with the dWOLS balancing weights, and acquire estimates ($\hat{\boldsymbol{\beta}}$ and $\hat{\boldsymbol{\psi}}$). Building on these estimates and the weights function (Equation \ref{neww}) which satisfies weights criterion (\ref{wtcri}), we can obtain new balancing weights for binary outcomes. Thus, the second step will utilize the logistic regression again with the new balancing weights to estimate the parameter of interest. 

dWGLM is doubly robust against misspecification of either the treatment or the treatment-free model.  If we misspecify the treatment model but correctly specify the treatment-free model, the estimator of blip parameters will be consistent. Alternatively, if the treatment-free model is misspecified, but we employ the balancing weights that are derived from a correct treatment model,  the approximate consistency of the blip parameters will also be ensured. In addition, we note that the blip parameters are only meaningful if the blip model is correctly defined.  To specify the optimal treatment strategy, we need to correctly specify the blip model. For continuous outcomes, Wallace et al. \cite{wallace2017model} develop methods for assessing the blip model specification, and similar problems for binary outcomes can be further investigated.


\section{Simulation}
We now demonstrate the implementation and double robustness of dWGLM via two simulation studies that address problems in both single-stage decision and multi-stage decision settings. In the single-stage setting (Study 1), we consider four different scenarios to verify the double robustness property of our method. In each scenario, we also consider two different link functions and compare out dWGLM with proposed new weights with two alternatives: Q-learning and GLM with standard "absolute value" weights. To test the robust estimation ability of our methods, in Study 2, with its multi-stage decision settings, we examine two different data-generating processes that can be employed in different real situations. One (Study 2a) is analogous to Wallace and Moodie \cite{wallace2015doubly}'s two-stage setting, while the other (Study 2b), which follows Moodie et al. \cite{moodie2012q}'s setting, distinguishes between the components that are tailoring variables and those that are predictive variables including potential confounders. In each of Study 2a and 2b, we also test different misspecification cases to demonstrate the double robustness of dWGLM. 

\subsection{Single-stage decision for binary outcomes}

\begin{figure}[H]\centering
\includegraphics[width=3.2in]{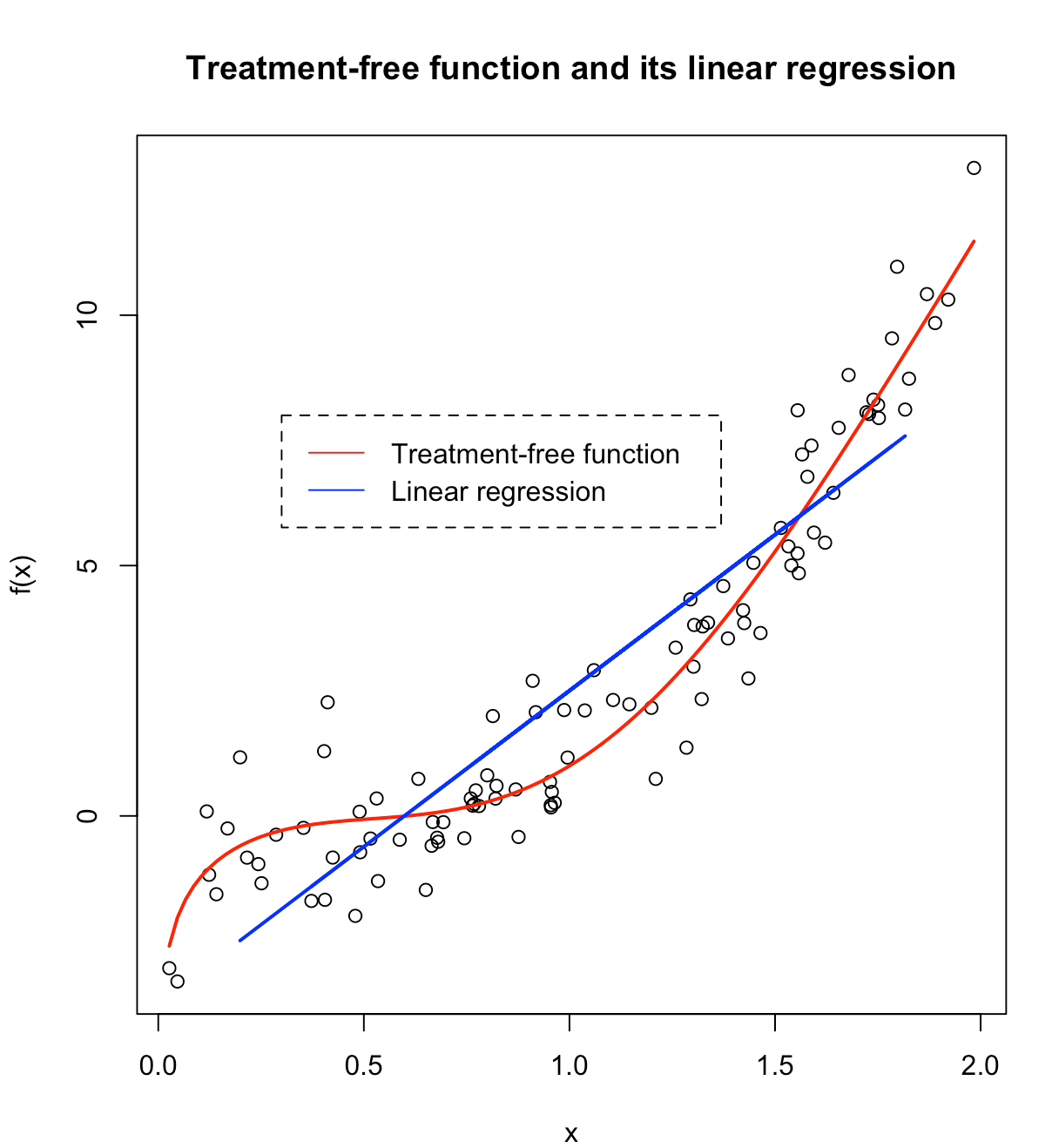} 
\caption{Nonlinear treatment-free function and its linear approximation}\label{treatment-freePlot}
\end{figure}

Our first simulations (\emph{Study 1}) consider the case of a single-stage treatment decision with binary outcomes. The data-generating process is as follows. Patient information: $X_i \sim U(0,2)$, where subscript $i$ indicates patient-level data; treatment $\mathbb{P}(A_i = 1 \mid X_i) = expit[ -2X_{i} + sin(X_{i}) + X_i^2 ]$, where $expit(x) = [1 + exp(-x)]^{-1}$; outcome $g[\mathbb{P}(Y_i = 1)]  = X_i + log(|X_i|) + cos(\pi X_i)+ X_i^3 + A_i(\psi_0+ \psi_1X_i)$, where both probit ($g(x)= \Phi^{-1}(x)$) and logit ($g(x) = x/(1-x)$) links are considered. Note that the treatment-free function is set as a nonlinear function that $f(x) = x + log(|x|) + cos(\pi x)+ x^3$, and its plot against $x$ is shown in Figure \ref{treatment-freePlot}. The blip function is set in the form $\gamma(x, a; \boldsymbol{\psi}) = a(\psi_0 + \psi_1x)$ with $\psi_0 = -1, \psi_1=2$, so that the optimal treatment is given by $a^{opt} = \mathbb{I}(\psi_0 + \psi_1x > 0)$ (or $a^{opt} = \mathbb{I}(x > 0.5)$). Our interest is then in estimating the blip parameters $\psi_0, \psi_1$. In this study, we consider three estimation methods. In \emph{Method 0} (Q-learning in Moodie and Krakow \cite{moodie2020precision}) we propose GLM with no weights for binary outcomes. In \emph{Method 1} (GLM with standard dWOLS weights), we consider GLM but with the original dWOLS "absolute value" weights (e.g., $w = |a - \mathbb{E}(A|\boldsymbol{x})|$). Finally, in \emph{Method 2}, which is our proposed method dWGLM, we consider GLM with the proposed weights (\ref{neww}), constructed from the standard dWOLS weights and the estimates from the model of (\ref{model3}).

For each simulation, we conduct analyses in the following four scenarios: 1) both the treatment and treatment-free models are mis-specified; 2) the treatment-free model is mis-specified but the treatment model is correctly specified; 3) the treatment model is mis-specified but the treatment-free model is correctly specified; and 4) both the treatment and treatment-free models are correctly specified. Model mis-specification is implemented via the omission of non-linear terms in the treatment and treatment-free models.


\begin{figure}[!]\centering
    \includegraphics[scale=0.45]{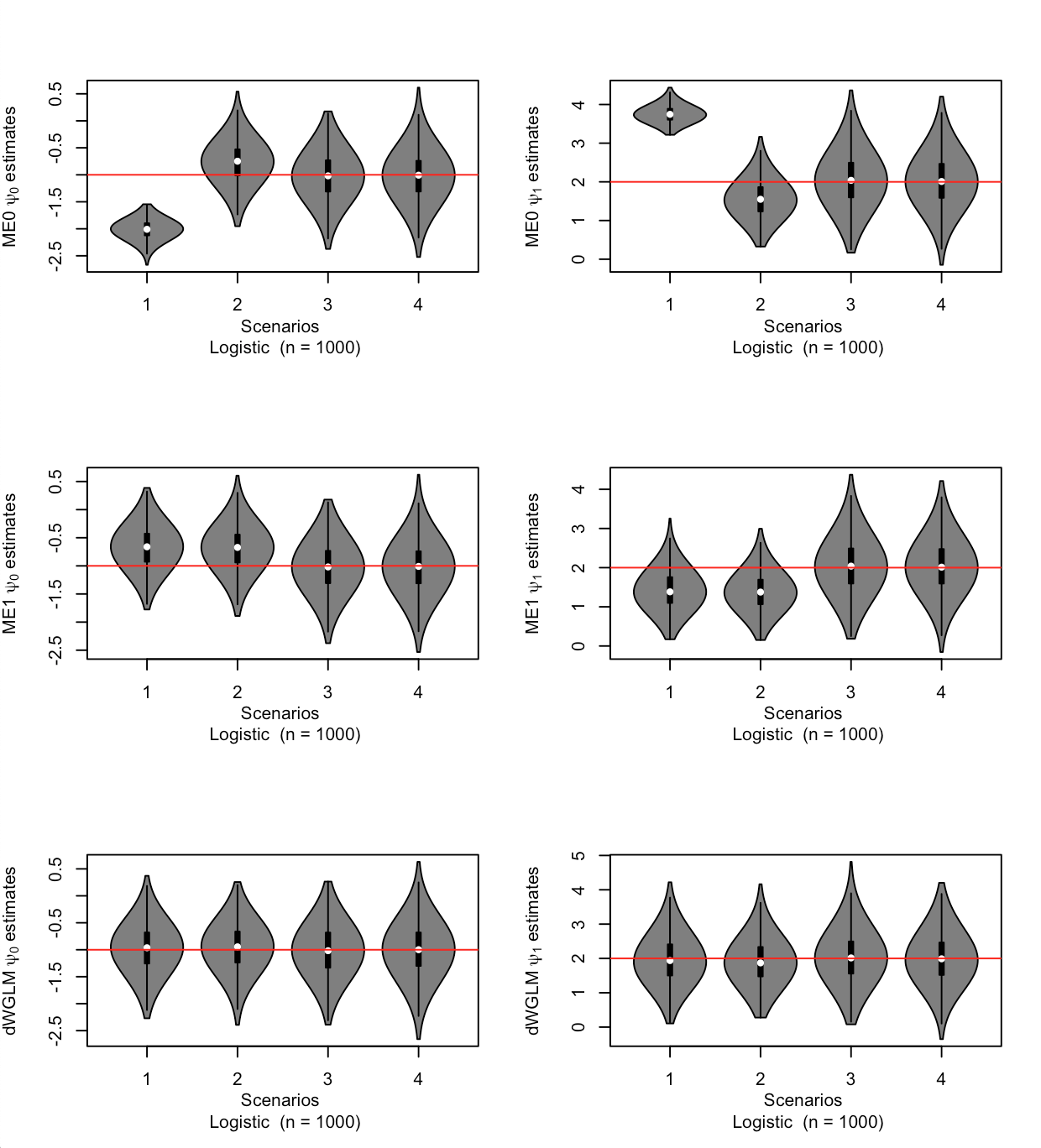} 
	\caption{Blip function parameter estimates via Method 0 (M0 Q-learning, top row), Method 1 (M1 GLM model with standard dWOLS weights, middle row) and Method 2 (dWGLM, bottom row) with logit link when neither model (scenario 1), treatment model only (scenario 2), treatment-free model only (scenario 3), or both models are correctly specified (scenario 4).}\label{single}
\end{figure}

Our simulation demonstrates the expected results as shown in Figure 2, which presents the results of the GLM with logit link, and those of the GLM with probit link appears a similar pattern. In the first two scenarios, where the treatment-free model is incorrectly specified, both Methods 0 and 1 provide biased estimators of blip function parameters. However, Method 2, the proposed dWGLM method with new balancing weights (\ref{neww}) offers blip estimators that are close to unbiased (and therefore likely to be close to consistent) in Scenario 2 and blip estimators with a small bias in Scenario 1. For the last two scenarios (Scenarios 3 and 4), because the treatment-free models were correctly specified, all of these three methods perform well in estimating the blip parameters; that is, they all provide close to consistent blip function parameter estimators. We note that, for dWGLM, comparing Scenario 1 with Scenario 2 (or Scenario 3 with Scenario 4), we observe a gain in efficiency due to the correct specification of the treatment model. 

\subsection{Two-stage decision for binary outcomes}
\begin{figure}[H]\centering
	\begin{tikzpicture}
    \node[state] (x_1) at (0,0) {$X_{1}$};
    \node[state] (x_2) [right =of x_1] {$X_{2}$};
    \node[state] (a_1) [below right =of x_1,xshift= -0.8cm,yshift=0.3cm] {$A_1$};
    \node[state] (y) [right =of x_2] {$Y$};
    \node[state] (a_2) [below right =of x_2,xshift= -0.8cm,yshift=0.3cm] {$A_2$};
     \path (x_1) edge (x_2);
    \path (x_2) edge (y);
    \path (x_1) edge[bend left=50] (y);
     \path (x_1) edge (a_1);
    \path (x_2) edge (a_2);
    \path (a_2) edge (y);
    \path (a_1) edge (y);
    \end{tikzpicture}
\caption{Directed Acyclic Graphs (DAG) of Simulations 3.2 (\emph{Study 2a}). }\label{fig:2}
\end{figure}
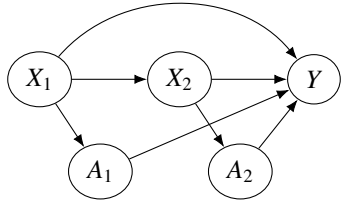

Our second set of studies will demonstrate the implementation of our strategies in simulated datasets for a two-stage treatment decision process. In \emph{Study2a}, we consider the outcome model of form: $logit[\mathbb{P}(Y = 1)] = logit[\mathbb{P}(Y^{opt} = 1)] - \sum_{j=1}^{K}\left[\gamma_{j}\left(\boldsymbol{h}_{j}, a_{j}^{o p t}\right)-\gamma_{j}\left(\boldsymbol{h}_{j}, a_{j}\right)\right]$, and examine the double robustness of the proposed dWGLM in the two-stage ($K = 2$) decision problem.   A causal diagram of this two-stage decision is shown in Figure \ref{fig:2}. Writing the column vector $\boldsymbol{\psi}_j = (\psi_{0j}, \psi_{1j})^{\top}$, the data-generating process is as follows.
\begin{itemize}
    \item Patient information: $X_{1} \sim N(2,1),$ $X_{2} \sim N(1+0.5X_1,2);$
    \item Treatment: $\mathbb{P}(A_1 = 1 \mid x_1 ) = expit[ -5 + x_{1} + x_{1}^2  ],$ $\mathbb{P}(A_2 = 1 \mid x_2 ) = expit[ -2.5x_{2} + x_{2}^2 + sin(x_{2})];$
    \item Blip functions: $\gamma_{j}\left(\boldsymbol{h}^{\psi}_{j} ; \boldsymbol{\psi}_{j}\right)=a_{j} \boldsymbol{\psi}^{\top}_{j}\boldsymbol{x}^{\psi}_{j},$ with $\boldsymbol{x}^{\psi}_{j} = (1, x_j)^{\top}$ , $\psi_{0j} = -2$ and $\psi_{1j} = -1$ for $j = 1,2.$ The regret function is thus $\mu_{j}\left(\boldsymbol{h}^{\psi}_{j} ; \boldsymbol{\psi}_{j}\right)=\left(a_{j}^{opt}-a_{j}\right)  \boldsymbol{\psi}^{\top}_{j}\boldsymbol{x}^{\psi}_{j},$ where $a_{j}^{opt} = \mathbb{I}(\boldsymbol{\psi}^{\top}_{j}\boldsymbol{x}^{\psi}_{j} > 0);$
    \item Outcome: $logit[\mathbb{P}(Y = 1)] = logit[\mathbb{P}(Y^{opt} = 1)] - \mu_{1}\left(\boldsymbol{h}^{\psi}_{1} ; \boldsymbol{\psi}_{1}\right) - \mu_{2}\left(\boldsymbol{h}^{\psi}_{2} ; \boldsymbol{\psi}_{2}\right),$ where $logit[\mathbb{P}(Y^{opt} = 1)]  =  x_{1} + log(|x_{1}|) + cos(\pi x_{1}).$
\end{itemize}

In this two-stage decision problem, to evaluate the double robustness property of our dWGLM approach, we consider various forms of model misspecification. In particular, we emphasize the following two cases: 1) the treatment-free models were misspecified for both stages by only considering linear terms, but the treatment models were specified correctly; 2) the treatment-free model was misspecified for the second stage, but the treatment model was specified correctly; in contrast, the treatment model was misspecified for the first stage, yet the treatment-free model was correctly identified.

\begin{figure}[h]\centering
	\begin{subfigure}[t]{.495\textwidth}
    \includegraphics[scale=0.33]{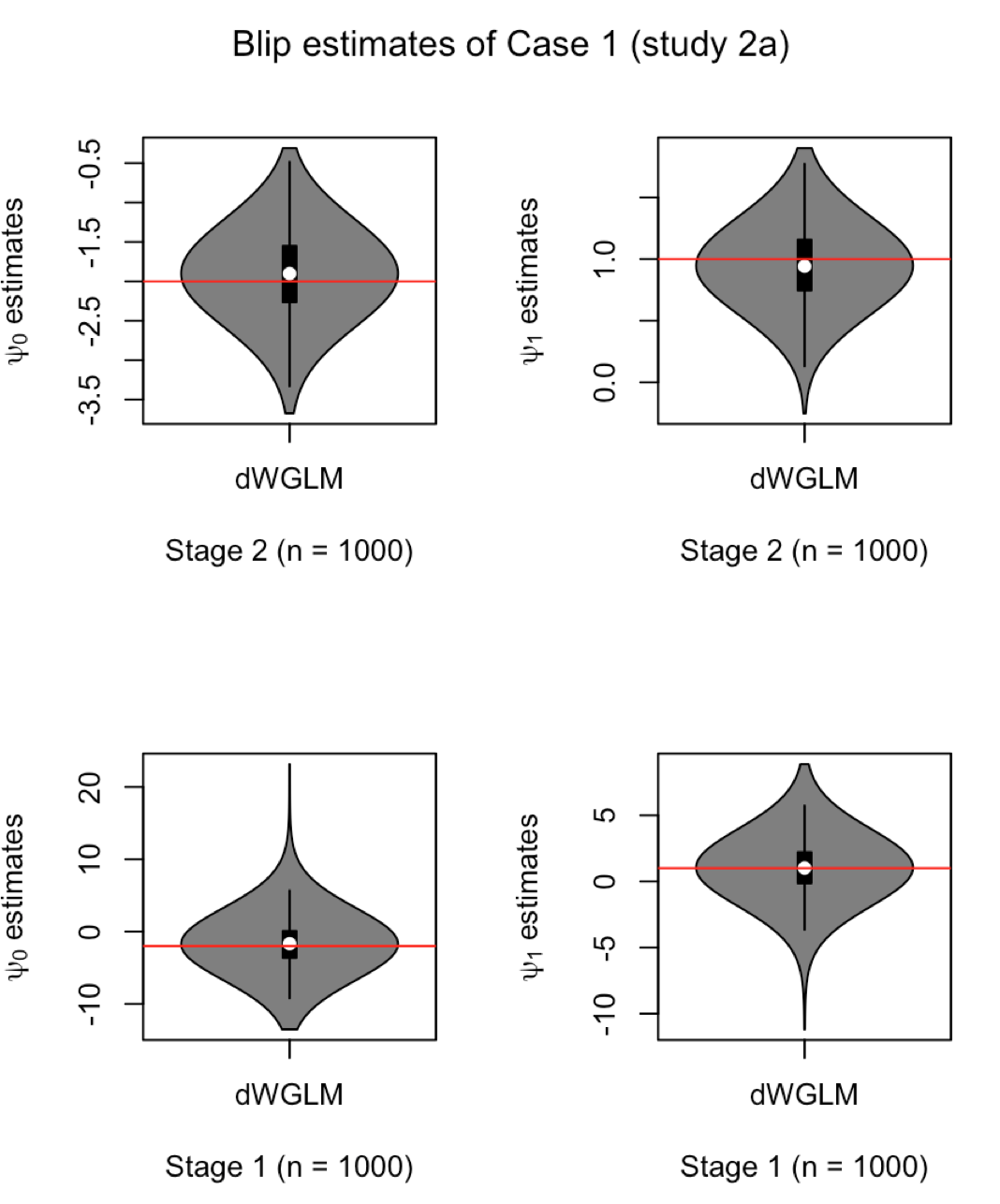} 
		\caption{Case 1}\label{fig:1a}		
	\end{subfigure}
	 \hfill
	\begin{subfigure}[t]{.495\textwidth}
    \includegraphics[scale=0.33]{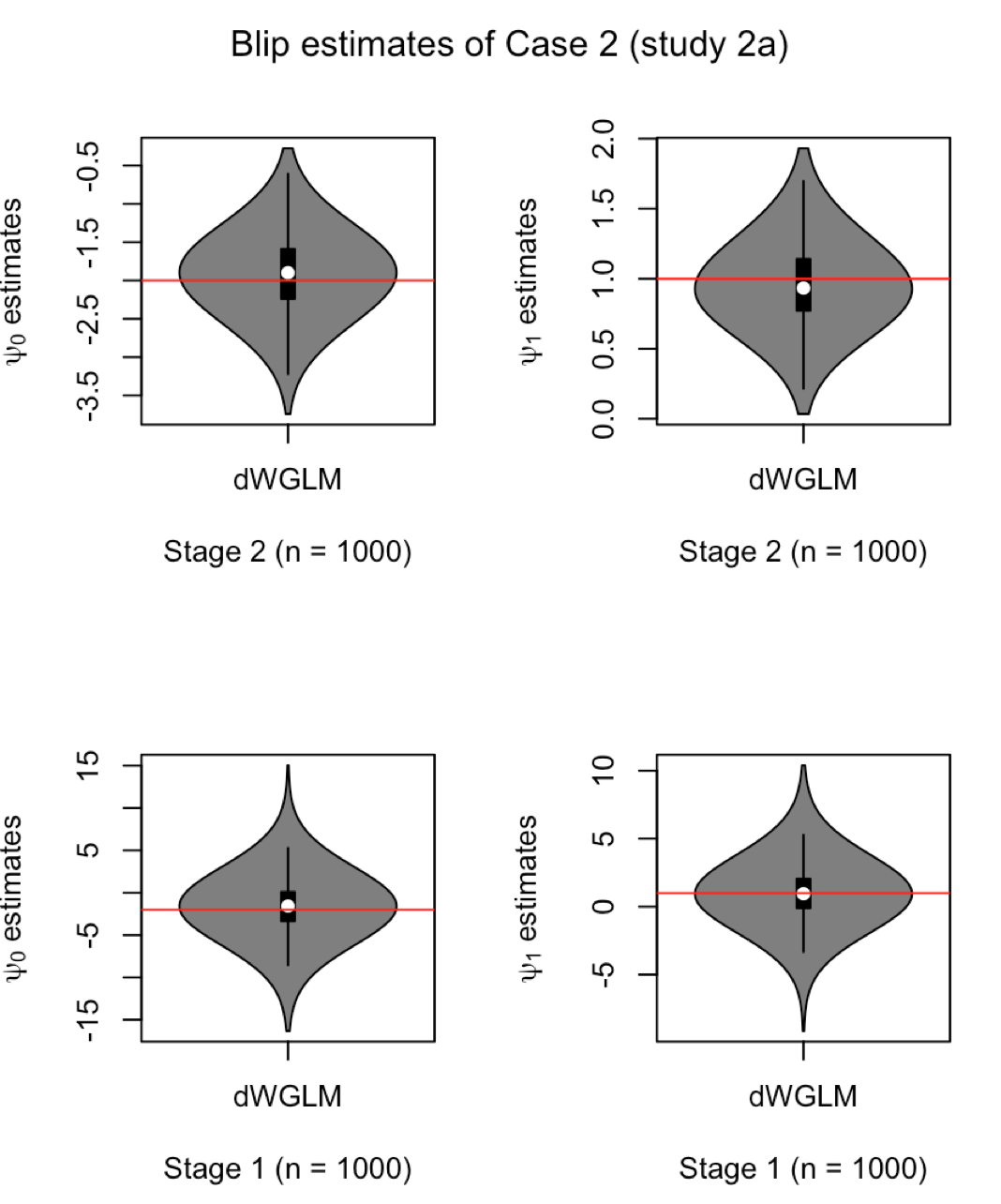} 
	\caption{Case 2}\label{fig:1b}
	\end{subfigure}
	\caption{Two-stage decision simulation Study 2a. The top row is the blip parameter estimates via dWGLM for Stage 2, and the bottom is the blip parameter estimates for Stage 1. The left two columns (a) and the right two columns (b) are estimates from Case 1 and 2, respectively.}\label{fig:3}
\end{figure}

For the two-stage binary-outcome problem where the binary outcome was generated by the model of the form $logit\left[\mathbb{P}\left(Y = 1\right)\right] = logit\left[\mathbb{P}\left(Y^{opt} = 1\right)\right] - \sum_{j=1}^{K}\left[\mu_{j}\left(\boldsymbol{h}_{j}, a_{j}\right)\right]$, the simulation results are as expected. The blip function parameter estimates are shown in Figure \ref{fig:3}. For Case 1, in both Stage 1 and Stage 2, the treatment-free models are incorrectly specified, but the treatment models are all correctly specified. The blip parameter estimates from both stages of Case 1 are plotted in Figure \ref{fig:1a}. The top row shows the blip parameter estimates via dWGLM for Stage 2, and the bottom gives the blip parameter estimates for Stage 1. Both stages’ blip parameter estimates appear to be consistent. 
In Case 2, where only the treatment model is correctly specified in Stage 2 and only the treatment-free model is correctly specified in Stage 1, our results show that the blip parameters (plotted in Figure \ref{fig:1b}) are also consistently estimated. Therefore, these results are as expected: the blip estimators appear consistent, and the double robustness of dWGLM in this study is verified.

\begin{figure}[H]\centering
	\begin{tikzpicture}
    \node[state] (a_1) at (0,0) {$A_{1}$};
    \node[state] (o_2) [right =of a_1] {$O_{2}$};
    \node[state] (a_2) [right =of o_2] {$A_{2}$};
    \node[state] (o_1) [left =of a_1] {$O_{1}$};
    \node[state] (x_1) [below right =of o_1,xshift= -0.8cm,yshift=0.1cm] {$X_1$};
     \node[state] (x_2) [below right =of o_2,xshift= -0.8cm,yshift=0.1cm] {$X_2$};
    \node[state] (y) [right =of a_2] {$Y$};
    \path (a_1) edge node[below, el]{$\delta_2$} (o_2);
    \path (a_2) edge (y);
    \path (x_1) edge (x_2);
    \path (x_1) edge node[above, el]{$\alpha_{11}$} (a_1);
    \path (x_2) edge node[above, el]{$\alpha_{12}$} (a_2);
    \path (x_2) edge (y);
    \path (x_1) edge[bend right=50] (y);
    \path (o_2) edge[bend left=25] (y);
    \path (a_1) edge[bend left=35] (y);
    \path (o_1) edge[bend left=50] (y);
    \path (o_1) edge[bend left=45] node[below, el]{$\delta_1$} (o_2);
    \end{tikzpicture}
\caption{Directed Acyclic Graphs (DAG) of Simulations 3.2 (\emph{Study 2b}). }\label{fig:4}
\end{figure}
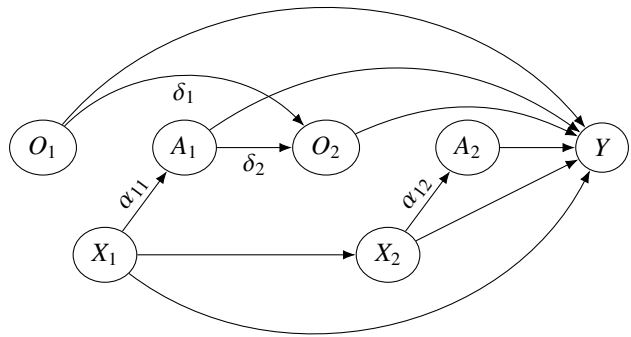

In \emph{Study2b}, motivated by real observational data where treatment assignment is confounded by covariates,  and based on the simulation study in Moodie et al. \cite{moodie2012q}, we distinguish between tailoring variables and predictive variables that include potential confounders, where the tailoring variables are denoted as $O_j$, and the predictive variables are denoted as $X_j$. The datasets feature some covariates recorded at each stage, where the second stage covariates are potentially affected by first stage variables (both treatment and non-treatment covariates). A causal diagram of this two-stage decision is shown in Figure \ref{fig:4}; therefore, the data generating process is as follows.

The covariates are $X_1 \sim N(3,1)$, $X_2 \sim N(-0.5 + 0.5X_1,1)$, and the treatment model is $\mathbb{P}(A_j = 1 \mid X_j) = expit(\alpha_{0j} + \alpha_{1j}X_j)$ for $j = 1,2$. The binary tailoring variables satisfy $\mathbb{P}(O_1 = 1) = 0.5$, and $\mathbb{P}(O_2 = 1 \mid O_1, A_1) = expit(\delta_1O_1 + \delta_2A_1)$. The outcome variable satisfies $$ \mathbb{P}(Y = 1 \mid X_1, O_1, A_1, X_2, O_2, A_2; \boldsymbol{\theta} ) = expit[m(x_1,x_2,o_1,o_2, a_1, a_2)],$$ where $m(x_1,x_2,o_1,o_2, a_1, a_2) = \theta_0 + \theta_1X_1 + \theta_2O_1 +\theta_3A_1 +\theta_4O_1A_1 +\theta_5X_2 +\theta_6 A_2 +\theta_7O_2A_2+\theta_8A_1A_2 + \varphi_1(X_1) + \varphi_2(X_2)$, and $\varphi_1$ and $\varphi_2$ may be non-linear functions such as $\varphi_1(X_1) = X_1^3$ and $\varphi_2(X_2) = log(|X_2|)$. We concentrated on the setting where $\boldsymbol{\theta} = (0, 1, 0, -0.5, -0.1, 1, 0.25,0.5,0.35)^{\top}$, $\boldsymbol{\delta} = (0.5, 0.6)^{\top}$, and $\boldsymbol{\alpha}_1 = (\alpha_{01}, \alpha_{11})^{\top} =  (-2.5, 1.25)^{\top},$ $\boldsymbol{\alpha}_2 = (\alpha_{02}, \alpha_{12})^{\top} =  (-0.5, 1.25)^{\top}$. We note that these choices of the parameters pertain to regular settings in the sense of Chakraborty et al. \cite{chakraborty2010inference}, but other choices that correspond to the non-regular settings can be further studied.

For the second stage, the true treatment-free function is $f_2 (x_1, o_1, a_1, x_2, o_2, a_2)= \theta_0 + \theta_1X_1 + \theta_2O_1 +\theta_3A_1 +\theta_4O_1A_1 +\theta_5X_2 + \varphi_1(X_1) + \varphi_2(X_2)$, and true blip function is $\gamma_2 (o_2, a_2, a_1) = \theta_6 A_2 +\theta_7O_2A_2+\theta_8A_1A_2$. Thus, for the second stage, the true blip parameters are $\boldsymbol{\psi}_2 = (\theta_6, \theta_7, \theta_8)^{\top} = (0.25, 0.5, 0.35)^{\top}.$ However, the true first stage decision rule parameters are more complicated because $O_2$ depends on $A_1.$ Building on the work of Moodie et al. \cite{moodie2014q}, in Appendix \ref{Appx.P2C1}, we derive the true first-stage decision rule parameters (i.e., $\boldsymbol{\psi}_1 = (\psi_{10}, \psi_{11})^{\top}$) as a function of the data-generating parameters. That is, for the true blip parameters $\boldsymbol{\psi}_1 = (\psi_{10}, \psi_{11})^{\top}$, we have the coefficient of $A_1$ as $$\psi_{10} = \theta_3 + |\phi_3|^{+} - |\phi_4|^{+} + k_3 \left(|\phi_1|^{+} - |\phi_3|^{+} \right) - k_4 \left(|\phi_2|^{+} - |\phi_4|^{+} \right),$$ and the coefficient of $O_1A_1$ as $$\psi_{11} = \theta_4 + (k_1-k_3) \left(|\phi_1|^{+} - |\phi_3|^{+} \right) - (k_2 - k_4) \left(|\phi_2|^{+} - |\phi_4|^{+} \right).$$

\begin{figure}[!]\centering
    \includegraphics[scale=0.35]{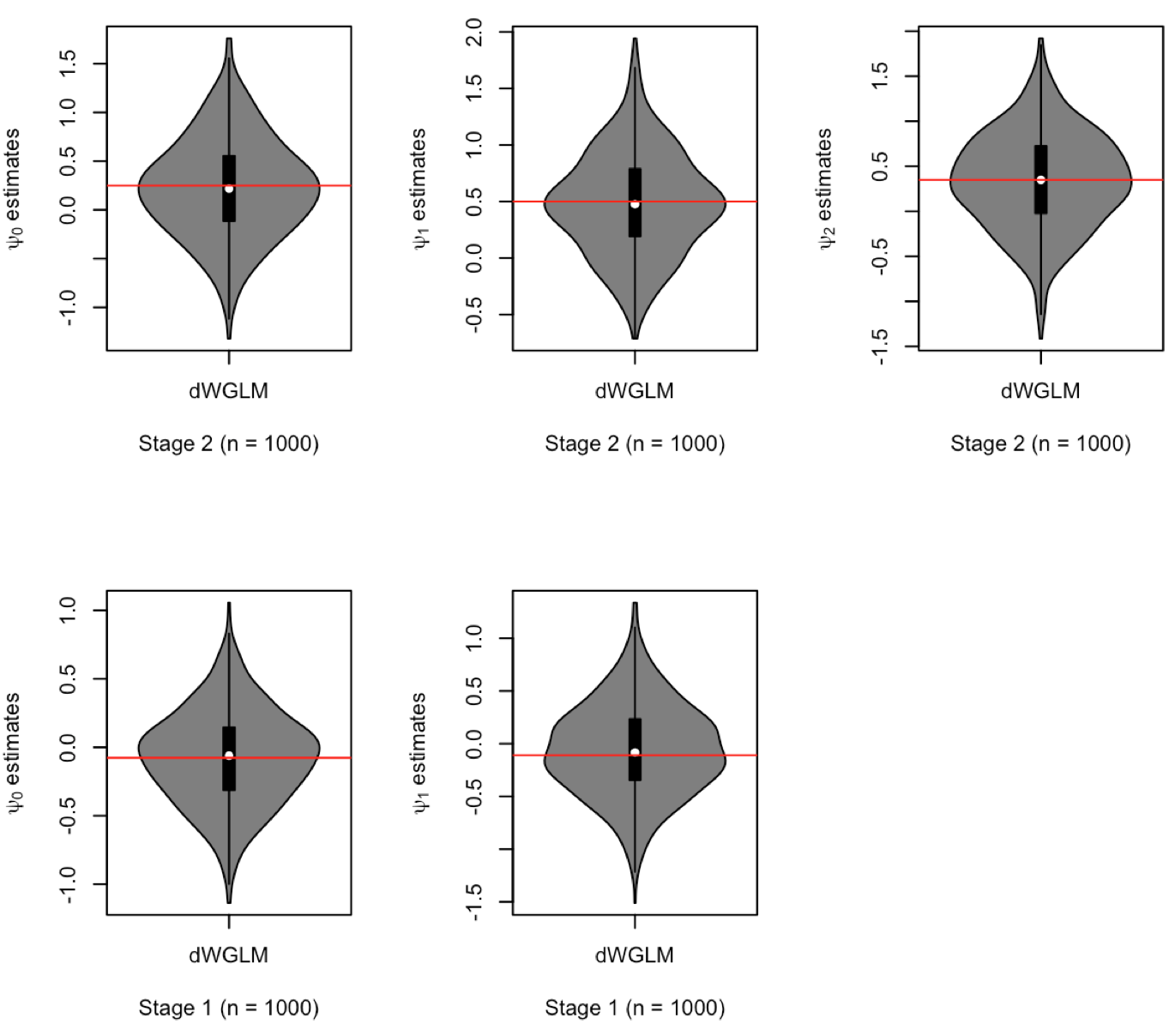} 
	\caption{Estimates of blip parameters in Study2b. Two-stage decision (Stage 2, top row) blip function parameter estimates via dWGLM with logit link (left to right columns are for $\psi_0$, $\psi_1$, and $\psi_2$, respectively) when only the treatment model is correctly specified.}\label{fig:5}
\end{figure}

Figure \ref{fig:5} shows the blip estimates from the simulation Study 2b, and we can conclude that the blip parameters appear to be consistently estimated; therefore, the results from Study 2b, with a parameterization that resembles that of a real dataset, are also as expected.

\section{Population Assessment of Tobacco and Health Study}
We now implement our approach using real data from a national longitudinal cohort study: the Population Assessment of Tobacco and Health (PATH) Study (\cite{hyland2017design}). The purpose of the  PATH Study is to collect data on tobacco use and how it affects the health of people in the United States. In analyzing the PATH data, our interest is in estimating the optimal DTR for each smoker, in terms of a sequence of use or non-use of e-cigarettes, to achieve smoking cessation.  This ongoing study collects data in \emph{waves}, starting from 2013, with each subsequent wave beginning approximately one year after the previous one. Altogether 53,178 participants, both adults and youth, and both smokers and non-smokers, constituted the first (baseline) wave, Wave 1.  We consider the subset of respondents who are smokers in Wave 1.   Using the first four waves of data, we formulate our analysis as a three-stage decision problem and define the $j^{th}$ stage ($j = 1, 2, 3$) to be the time from Wave $j$ up to but not including Wave $j+1$. The PATH Study spans a time of changes in the e-cigarette market: e-cigarettes were starting to see widespread use in the United States at the start of the study, and the prevalence of e-cigarettes grew considerably between Waves 1 and 2, and between Waves 2 and 3. Later, Wave 4 coincided with the emergence of products that saw popular use particularly among younger people (\cite{huang2019vaping}). A growing body of literature suggests that e-cigarettes (vaping) can be a useful smoking-cessation aid  (\cite{villanti2018we}; \cite{hajek2019randomized}); thus, we set the treatment variable as the use of e-cigarettes by a cigarette smoker. Due to the long participant-follow-up of approximately one year, we define e-cigarette use reported at the wave of the measured outcome as indicative of the pre-wave treatment (i.e., e-cigarette use).  In addition, the e-cigarettes usage variable is determined by participants being asked during the study whether they “now use e-cigarettes (a) Every day (b) Some days (c) Not at all.” If participants answer either “Every day” or “Some days,” then they are deemed to use e-cigarettes and are coded $A = 1$; otherwise, if they respond “Not at all,” they are coded $A = 0$. The binary outcome in our analysis is the indicator of smoking cessation (of traditional cigarettes only) or not based on the question “Do you now smoke cigarettes (a) Every day (b) Some days (c) Not at all?” in the study. If participants respond (c), then their binary outcomes are coded as $Y=1$; if they respond (a) or (b), then their binary outcomes are coded as $Y=0$. 

Building on previous PATH analyses such as that in Benmarhnia et al. \cite{benmarhnia2018can}, for the $j^{th}$ stage, we select the Wave $j$ variables age (``less than 35” or ``35+”), education, sex, non-Hispanic, race and ``plan to quit'', denoted, respectively, as the  covariates $\boldsymbol{x}_j^{\beta} = (x_{j1}, x_{j2}, x_{3},x_{4}, x_{5}, x_{j6})^{\top}$ in the treatment-free model. We note that the questionnaire will no longer ask the question regarding “plan to quit”  to participants who have already quit smoking; thus, we assign the value of that question to 1 (i.e., they have a plan to quit smoking) for those participants. In addition, the tailoring variables that are related to the efficacy of the treatment should be selected from a set of moderator variables (\cite{almirall2014introduction}). Building on  previous work of studying  moderators in the relationships of prior wave predictors of quitting smoking, we select at each stage the variables age and ``plan to quit'' as tailoring variables, i.e., $\boldsymbol{x}_j^{\psi} = (x_{j1}, x_{j6})^{\top}$ (\cite{le2020predictive}). The covariates in the treatment propensity models are chosen based on the work of Benmarhnia et al. \cite{benmarhnia2018can}, and $\boldsymbol{x}_j^{\alpha}= \boldsymbol{x}_j^{\beta} = (x_{j1}, x_{j2}, x_{3},x_{4}, x_{5}, x_{j6})^{\top}$. Therefore, in estimation, the blip model is set up as $\gamma\left( \boldsymbol{x}_j^{\psi}; \boldsymbol{\psi}_j \right)=a_{j+1}\left(\psi_{j0}+ \boldsymbol{\psi}_{j1}^{\top} \boldsymbol{x}_j^{\psi}\right),$ and the treatment-free model as $f^{}_j(\boldsymbol{x}_{j}^{\beta}; \boldsymbol{\beta}_j) = \beta_{j0} + \boldsymbol{\beta}_{j1}^{\top} \boldsymbol{x}_{j}^{\beta}$. 
Four sets of analyses corresponding to those carried out in the simulations of Section 3.1 are conducted. At each stage of Method 0 (M0), i.e., the Q-learning approach, a logistic regression is implemented. Method 1 is similar to M0 but uses the dWOLS balancing weights in each logistic regression estimation. Then, Method 2 (M2), our proposed doubly-robust method, uses the logistic regression model with the new weights based on the equation (\ref{neww}). To perform a sensitivity analysis for the link function, we also consider Method 3 (M3), which is analogous to M2 as it includes two-step robust estimation in each stage, but which uses the probit link function. We note that only M0 does not use any balancing weights in the estimation process, but M1, M2, and M3 do use weights for the purpose of balancing. 

\begin{table}[!]\centering
\setlength{\tabcolsep}{5pt}
\caption{Analysis and optimal DTRs of PATH data. Optimal DTRs are indicator functions of $\boldsymbol{\psi}$ estimates. }\label{t:path} 
\begin{tabular}{cllllll}
\hline
\multirow{2}{*}{Wave} & \multirow{2}{*}{Estimates $\boldsymbol{\hat{\psi}}$} & \multicolumn{4}{c}{ Methods }                               \\ 
                  &                   & \multicolumn{1}{c}{M0}  & \multicolumn{1}{c}{M1}  & \multicolumn{1}{c}{M2}  &\multicolumn{1}{c}{ M3} \\ \hline
\multirow{3}{*}{$1 \sim 2$} & \multicolumn{1}{c}{$\hat{\psi}_0$ }               & \multicolumn{1}{c}{0.0188} & \multicolumn{1}{c}{0.0236} & \multicolumn{1}{c}{-0.0013} & \multicolumn{1}{c}{-0.0041} \\ 
                  & \multicolumn{1}{c}{$\hat{\psi}_1$ }             & \multicolumn{1}{c}{-0.0841} & \multicolumn{1}{c}{-0.0710} &\multicolumn{1}{c}{-0.0250} & \multicolumn{1}{c}{-0.0055}  \\ 
                   & \multicolumn{1}{c}{$\hat{\psi}_2$ }            & \multicolumn{1}{c}{0.1142} & \multicolumn{1}{c}{0.0428} & \multicolumn{1}{c}{-0.0219} & \multicolumn{1}{c}{0.0014} \\ \hline
\multirow{3}{*}{$2 \sim 3$} & \multicolumn{1}{c}{$\hat{\psi}_0$  }              & \multicolumn{1}{c}{0.0602} & \multicolumn{1}{c}{0.0975} & \multicolumn{1}{c}{0.1244} & \multicolumn{1}{c}{0.0380} \\ 
                  & \multicolumn{1}{c}{$\hat{\psi}_1$  }            & \multicolumn{1}{c}{-0.0714} & \multicolumn{1}{c}{-0.1114} &\multicolumn{1}{c}{-0.1229} & \multicolumn{1}{c}{ -0.0345} \\ 
                   & \multicolumn{1}{c}{$\hat{\psi}_2$ }            & \multicolumn{1}{c}{0.0315} & \multicolumn{1}{c}{0.0693} & \multicolumn{1}{c}{0.1198} & \multicolumn{1}{c}{0.0150} \\ \hline
\multirow{3}{*}{$3 \sim 4$} & \multicolumn{1}{c}{$\hat{\psi}_0$}                & \multicolumn{1}{c}{-0.0469} & \multicolumn{1}{c}{0.0035} & \multicolumn{1}{c}{-0.1553} & \multicolumn{1}{c}{0.0088} \\ 
                  & \multicolumn{1}{c}{$\hat{\psi}_1$  }            & \multicolumn{1}{c}{0.2535} & \multicolumn{1}{c}{0.2478} &\multicolumn{1}{c}{0.4581} &  \multicolumn{1}{c}{0.1547} \\ 
                   & \multicolumn{1}{c}{$\hat{\psi}_2$  }           & \multicolumn{1}{c}{0.2937} & \multicolumn{1}{c}{0.2621} & \multicolumn{1}{c}{0.1745} & \multicolumn{1}{c}{0.1261} \\ \hline

\end{tabular}
\end{table}

Our proposed new weights (equation \ref{neww}) are built to provide an unbiased estimation of the blip parameters through estimating weighted population-level estimating equations. In the PATH data analysis process, we also employ the sampling design weights in each stage. Regarding the combination of model and sampling design, the sample estimating functions are unbiased with respect to the sample design of the population-level estimating functions; thus, the estimators of the blip function parameters are model-design consistent (\cite{lumley2004analysis}).

As shown in Section 2, our use of balancing weights is to ensure consistent blip estimators.  M2 and M3, which employ the proposed balancing weights, are expected to provide consistent blip estimators, but M1, which uses standard dWOLS weights, and M0, which does not use any balancing weights, are not. The blip parameter estimates from Methods 0, 1, 2, and 3 are summarized in Table \ref{t:path}. The results of M0 and M1 are similar, especially $\hat{\psi}_0$ and $\hat{\psi}_1$ in Stage 1 (Wave $1 \sim 2$), and $\hat{\psi}_1$ and $\hat{\psi}_2$ in Stage 3 ($3 \sim 4$). Both M2 and M3 employ the proposed balancing weights, but they use different link functions and return different estimates. The difference between estimates from M1 and M2 shows the difference in results obtained with our method when employing two different weights: the original dWOLS weights that are for the continuous outcome model and our proposed balancing weights that are for the binary outcomes. We also see differences between M2 (using the logit link) and M3 (using the probit link). Compared with the estimates of M2, those of M3 are attenuated; this is expected because the standard normal distribution has a lighter tail than the logistic distribution. For example, given a certain probability larger than 0.5, the value of the inverse of the logistic function is greater than that of the inverse of the standard normal distribution; thus the coefficient estimates obtained with the probit link function will tend to be smaller compared with those obtained with the logit link function if the covariates are the same. Although the estimates of M2 and M3 are different, their patterns that decide the treatment recommendations are similar. 

Building on the blip parameter estimates, the corresponding optimal treatment regime will be $\hat{a}_j^{opt} = \mathbb{I}(\boldsymbol{\hat{\psi}}^{\top}_j\boldsymbol{x}_j^{\psi} > 0)$, for $j = 1,2,3$; for instance, for Wave $1 \sim 2$, Method 2 (i.e., dWGLM) outputs $\mathbb{I}[-0.0013 -0.0250x_{11} -0.0219x_{16} > 0]$; for Wave $2 \sim 3$, Method 2 outputs $\mathbb{I}[0.1244 -0.1229x_{11} +0.1198x_{16} > 0]$, and for Wave $3 \sim 4$, it outputs $\mathbb{I}[-0.1553 +0.4581x_{11} + 0.1745x_{16} > 0]$. These results from Method 2 can be interpreted as the following treatment recommendations about the use of e-cigarettes. In the first stage, do not use e-cigarettes. In the second stage, use e-cigarettes. In the last stage, if a smoker's age is less than 35 and he or she has no plan to quit, do not use e-cigarettes; otherwise, use e-cigarettes.  Finally, we note that, through the three-stage treatment decision analysis of the PATH data, our intention is mainly to illustrate that our dWGLM can be applied in practice, but not to put forward the results as authentic recommendations for a treatment strategy.

\section{Conclusion and Discussion}
Dynamic treatment regimes are a mechanism by which treatment decisions are made based upon individual-level information,  used in optimizing long-term expected outcomes. Many approaches for optimal DTR estimation are limited to continuous outcomes. The few used to address binary outcomes are limited in their robustness to model mis-specification or complexity of implementation. Our dWGLM method, motivated by its continuous-outcome predecessor dWOLS, provides double robustness to model misspecification while being comparatively easy to implement. Our method can be viewed as a series of weighted GLM analyses. Meanwhile, to make optimal sequential decisions, some care in constructing stage-specific pseudo-outcomes is also needed. We offer a new balancing weight criterion to overcome the misspecification of treatment-free models, and the method for each decision stage involves just a two-step logistic regression.

Our dWGLM approach is doubly robust for estimating the parameter of interest, a property demonstrated via simulation. It is important to acknowledge that our approach relies on the suitability of the local linear approximation to the inverse link function ($g^{-1}$). We utilize Taylor expansion of $g^{-1}$  about $\boldsymbol{\beta}^{\top}X$ evaluated at $f(X)$ (Equation \ref{eq:A.8} in Appendix \ref{Appx.P2A1}) in the proof of Theorem \ref{P2thm1}, and the error term is close to zero if a linear predictor tends to vary in an interval where $g^{-1}$ is close to linear. At the end of Appendix \ref{Appx.P2A1}, we discuss the accuracy of approximation that can be decided by the range of covariates in our method and the inverse link function. Therefore, a possible extension would be to conduct sensitivity analyses to study different link functions as well as treatment-free functions in GLM. 

In future work, we note that some machine learning (ML) methods can be employed in our dWGLM analytical framework.  For example, tree-based methods (e.g., Bayesian additive regression trees, \cite{chipman2010bart}) are commonly used in estimating the treatment model, and some ensemble methods (e.g., Super Learner, \cite{van2007super}) can be used for last stage estimation to provide a more accurate prediction of $\mathbb{P}(Y = 1 \mid \boldsymbol{h}_K, a_K)$, and thus to produce accurate pseudo outcome prediction. For another example, Moodie et al. \cite{moodie2012q} employs a generalized additive model in Q-learning. The ML methods, of course, are chosen based on the purpose of the estimation. For the last stage outcome model, the ML method is required for precise prediction; however, for the treatment model, correct modeling of the data generating mechanism is not necessary, but rather all confounders must be included to correctly model the impact of the treatment (\cite{ertefaie2012estimation}). Further, due to regularization and overfitting, the "prediction-focused" ML estimators may be biased (\cite{chernozhukov2018sorted}); thus, orthogonalization and data splitting should be carefully investigated to control the regularization and the overfitting bias. Therefore, one important extension to our work is to employ different ML models in the corresponding process to produce accurate optimal treatment regimes.

\begin{ack}
This work has been supported by the Ontario Institute for Cancer Research (OICR) BTI Studentship Award through funding provided by the Government of Ontario, by a CIHR Project Grant to M.P. Wallace, and by a Discovery Grant to M. E. Thompson  (RGPIN-2016-03688) from NSERC. \\
Special thanks are due to the use of  PATH data \cite{PATH}.
\end{ack}


\renewcommand\refname{Bibliography}
\bibliographystyle{apalike}
\bibliography{reference.bib}

\newpage
\begin{appendix}
\section{Proof of Theorem 1: Balancing property for binary outcomes}\label{Appx.P2A1}
This section presents the proof of Theorem 1. Building on the proof, the two-step estimation process of the proposed methods is also shown at the end. 

Let us consider the true model: $g(\mathbb{E}[Y |A,X]) = f(X^\beta) + \gamma(A,X^{\psi}; \boldsymbol{\psi})$, where $Y \in \{0, 1\}$, and $X \in \mathbb{R}^{p}$ is a vector of $p$ covariates; $X^{\beta} \in \mathbb{R}^{p^{\beta}}$ and $X^{\psi} \in \mathbb{R}^{p^{\psi}}$ are two (potentially identical) subsets of the
variables contained in $X$; $A \in \{0,1\}$ is the treatment assignment. Their realized values are $y$, $x$, $x^{\beta}$, $x^{\psi}$ and $a$, respectively, and we write the propensity score as
$\mathbb{P}(A = 1 \mid x) = \pi(x)$. Note that the link function that relates the  predictor to the expected value of the random variable $Y$ is denoted as $g(*)$.  We assume that $\gamma(A,X^{\psi};  \boldsymbol{\psi})$ is correctly specified, and set $\gamma(A,X^{\psi}; \boldsymbol{\psi}) =  \boldsymbol { \psi }^{\top}A X^{\psi}$. 

In the case that $f(X^\beta)$ is linear, as the estimation model assumes, let a latent continuous outcome defining the observed outcome be $Y_{c} = \boldsymbol { \beta }^{\top} X^{\beta} + \boldsymbol { \psi }^{\top} A X^{\psi} + \epsilon$, where the error term $\epsilon$ could be assumed to follow a logistic distribution conditional on the explanatory variables. This generates the standard logistic model. However, it is not necessary that  $\epsilon$  has a logistic distribution. It could have a standard normal distribution, yielding a probit model, or another reasonable distribution. The cumulative distribution function of $\epsilon$ is the inverse link function, i.e., $g^{-1}(*)$.  Define the binary outcome $Y$ as a dichotomization of the latent continuous outcome $Y_c$, such that $Y=\mathcal{I}\left(Y_{c} \geq 0\right) =\mathcal{I}\left(\epsilon \geq - \boldsymbol { \beta }^{\top} X^{\beta} - \boldsymbol { \psi }^{\top} A X^{\psi}\right)$, where $\mathcal{I}$ is the indicator function.

Thus $ \mathbb{P}(Y=1 \mid x)= \mathbb{P}\left(Y_{c} \geq 0 \mid x\right)=1- g^{-1}(- \boldsymbol { \beta }^{\top} X^{\beta} - \boldsymbol { \psi }^{\top} A X^{\psi})$.  For example, the logistic model is
$ \mathbb{P}(Y=1 \mid x)= \mathbb{P}\left(Y_{c} \geq 0 \mid x\right)=1-\frac{\exp \left(0- \boldsymbol { \beta }^{\top} X^{\beta} - \boldsymbol { \psi }^{\top} A X^{\psi}\right)}{1+\exp \left(0- \boldsymbol { \beta }^{\top} X^{\beta} - \boldsymbol { \psi }^{\top} A X^{\psi}\right)}=\frac{\exp \left( \boldsymbol { \beta }^{\top} X^{\beta} + \boldsymbol { \psi }^{\top} A X^{\psi}\right)}{1+\exp \left( \boldsymbol { \beta }^{\top} X^{\beta} + \boldsymbol { \psi }^{\top} A X^{\psi}\right)}$. Therefore, $$ \mathbb{P}(Y=y \mid x) = \frac{\exp \left[y( \boldsymbol { \beta }^{\top} X^{\beta} + \boldsymbol { \psi }^{\top} A X^{\psi})\right]}{1+\exp \left( \boldsymbol { \beta }^{\top} X^{\beta} + \boldsymbol { \psi }^{\top} A X^{\psi}\right)}    \text { for } y = 0,1, $$ and the log-likelihood for logistic
regression is 
$$
\begin{aligned} L(\boldsymbol{\beta}, \boldsymbol{\psi} \mid y, x) &=\log \prod_{i} \frac{\exp \left[y_{i} (\boldsymbol { \beta }^{\top} X_i^{\beta} + \boldsymbol { \psi }^{\top} A_i X_i^{\psi})\right]}{1+\exp \left(\boldsymbol { \beta }^{\top} X_i^{\beta} + \boldsymbol { \psi }^{\top} A_i X_i^{\psi}\right)} \\ &=\sum_{i: y_{i}=1} (\boldsymbol { \beta }^{\top} X_i^{\beta} + \boldsymbol { \psi }^{\top} A_i X_i^{\psi})-\sum_{i} \log \left(1+\exp \left(\boldsymbol { \beta }^{\top} X_i^{\beta} + \boldsymbol { \psi }^{\top} A_i X_i^{\psi}\right)\right). \end{aligned} $$
Thus, the score function system components are
\begin{equation}
    \sum_{i}\left(y_{i}-\frac{\exp \left(\boldsymbol { \beta }^{\top} X_i^{\beta} + \boldsymbol { \psi }^{\top} A_i X_i^{\psi}\right)}{1+\exp \left(\boldsymbol { \beta }^{\top} X_i^{\beta} + \boldsymbol { \psi }^{\top} A_i X_i^{\psi}\right)}\right) X_i^{\beta}
\end{equation}

and

\begin{equation}
    \sum_{i}\left(y_{i}-\frac{\exp \left(\boldsymbol { \beta }^{\top} X_i^{\beta} + \boldsymbol { \psi }^{\top} A_i X_i^{\psi}\right)}{1+\exp \left(\boldsymbol { \beta }^{\top} X_i^{\beta} + \boldsymbol { \psi }^{\top} A_i X_i^{\psi}\right)}\right) A_i X_i^{\psi}.
\end{equation}

Given posited outcome regression model $Q(X,A;\boldsymbol { \beta }, \boldsymbol { \psi }) = g^{-1}(\boldsymbol { \beta }^{\top} X^{\beta} + \boldsymbol { \psi }^{\top} A X^{\psi})$, the weighted GLM estimator for $(\boldsymbol { \beta }^{\top} , \boldsymbol { \psi }^{\top})^{\top}$ is obtained from solving the system of estimating equations: $\sum_i^nU_i (\hat{\boldsymbol { \beta }},\hat{ \boldsymbol {\psi}} ; A_i, X_i) = \boldsymbol{0}$, that is,

\begin{align}\label{eq1}
\sum_i^n \left(\begin{array}{c}X_i^{\beta} \\ A_iX_i^{\psi}\end{array}\right) w(A_i, X_i)\left[Y_i - g^{-1}(\boldsymbol { \psi }^{\top} A_i X_i^{\psi} + \boldsymbol { \beta }^{\top} X_i^{\beta})\right] =\sum_i^n\left[\begin{array}{c}U_{1i}(\boldsymbol { \beta }, \psi ; A_i, X_i) \\ U_{2i}(\boldsymbol { \beta }, \psi ; A_i, X_i)\end{array}\right]= \boldsymbol {0}
\end{align}

Based on the strong heredity principle (\cite{chipman1996bayesian}),  it is required that the treatment-free model must include the main effects for all covariates in the blip model, that is, the tailoring variables should be a subset of the predictive variables ($X^{\psi} \subseteq X^{\beta} $). For simplicity, and without loss of generality, we assume that $ X^{\psi} =X^{\beta} = X$, that is, the covariates in the treatment free and blip components are the same. Then, we assume that for each $i$ independently, $X_{i}$ for individuals are independent and identically distributed and are generated first, thereafter $A_i$ followed by $Y_i$. On the one hand, we consider

\begin{align}
\sum_i^n (U_{1i} - U_{2i})&=\sum_i^n (1-A_i)X_i w(A_i, X_i)\left[Y_i - g^{-1}(\boldsymbol { \psi }^{\top} A_i X_i^{\psi} + \boldsymbol { \beta }^{\top} X_i^{\beta})\right] \label{eq21}\\
& = \sum_i^n (1-A_i)X_i w(A_i, X_i)\left[Y_i - g^{-1}(  \boldsymbol { \beta }^{\top} X_i^{\beta})\right] = \boldsymbol{0} \label{eq22}
\end{align}

where the second equality follows because the only non-zero terms (in equation (\ref{eq21})) will be those for which $A_i = 0$. Thus, the left hand side of Equation (\ref{eq22}) (or $\sum_i^n (U_{1i} - U_{2i})$) does not depend on $\boldsymbol { \psi }$. We note that $\sum_i^n (U_{1i} - U_{2i}) = \boldsymbol{0}$ can be solved for $\boldsymbol {\beta }$, and its solution is denoted by $\hat{\boldsymbol{ \beta }}$.

The expectation of $\sum_i^n (U_{1i} - U_{2i})$ conditional on $(A_1, …. A_n)$ and $(X_1, ...X_n)$, that is, $ \sum_i^n (1-A_i)X_i w(A_i, X_i)\left[g^{-1}(f(X_i)) - g^{-1}(  \boldsymbol { \beta }^{\top} X_i^{\beta})\right]$, is not zero unless the true treatment-free model $f(X; \boldsymbol {\beta })$ is linear in $X$ with true coefficient $\boldsymbol {\beta }$. However, the expectation of $\sum_i^n (U_{1i} - U_{2i})$ conditional on $(X_1, ...X_n)$ is $ \sum_i^n (1- \pi(X_i))X_i w(0, X_i)\left[g^{-1}(f(X_i)) - g^{-1}(  \boldsymbol { \beta }^{\top} X_i^{\beta})\right]$, and if the (unconditional) expectation of $ \sum_i^n (1- \pi(X_i))X_i w(0, X_i)\left[g^{-1}(f(X_i)) - g^{-1}(  \boldsymbol { \beta }^{\top} X_i^{\beta})\right]$, i.e., $n\mathbb{E}\left[(1 - \pi(X))Xw(0, X)\left[g^{-1}(f(X)) - g^{-1}(  \boldsymbol { \beta }^{\top} X^{\beta})\right] \right]$, is $\boldsymbol{0}$ for $\boldsymbol {\beta} = \boldsymbol { {\beta}^{*}}$ and if $\boldsymbol { {\beta}^{*}}$ is unique, then, according to large sample theory, $\hat{\boldsymbol { \beta }} $ tends to $\boldsymbol { {\beta}^{*}}$ as $n \rightarrow \infty$ (\cite{white1982maximum}). Note that the uniqueness of $\boldsymbol { {\beta}^{*}}$ is shown in Appendix \ref{Appx.P2B1}.

On the other hand, if we consider 

\begin{align}
\sum_i^nU_{2i}= \sum_i^nA_iX_iw(A_i, X_i)\left[Y_i - g^{-1}(\boldsymbol { \psi }^{\top} A_i X_i^{}+\boldsymbol { \beta }^{\top} X_i^{})\right], 
\label{eq:A.8}
\end{align}

where the only non-zero terms are those for which $A_i = 1$, then $\hat{\boldsymbol{\psi}}$  can be solved in terms of $\hat{\boldsymbol {\beta }} ^ {}$ from $\sum_i^nU_{2i}=\boldsymbol{0}$. In order to show that $\hat{\boldsymbol{\psi}}$ is consistent, we would need to show that the expectation of 
$$\sum_i^nA_iX_iw(A_i, X_i)\left[g^{-1}(\boldsymbol { \psi }^{\top} A_i X_i^{}+\boldsymbol {{\beta }^{*}}^{\top} X_i^{}) - g^{-1}(\boldsymbol { \psi }^{\top} A_i X_i^{}+f(X_i))\right]$$ equals or is close to $\boldsymbol{0}$ for general $\boldsymbol { \psi }$.   If this is not the case, then the expectation of the equation $\sum_i^nU_{2i} = \boldsymbol{0}$  with $\boldsymbol{\beta} = \hat{\boldsymbol{\beta}}$ and $\boldsymbol{\psi} = \hat{\boldsymbol{\psi}}$ may approach the equation $\sum_i^nU_{2i} = \boldsymbol{0}$  with $\boldsymbol{\beta}$ set equal to $\boldsymbol { {\beta}^{*}}$ and $\boldsymbol{\psi}$ set equal to a similar limiting value $\boldsymbol { {\psi}^{*}}$ as $n \rightarrow \infty$.  The vector $\boldsymbol { {\psi}^{*}}$  will satisfy the condition that the expectation of  $\sum_i^nA_iX_iw(A_i, X_i)\left[g^{-1}(\boldsymbol { \psi^* }^{\top} A_i X_i^{}+\boldsymbol {{\beta }^{*}}^{\top} X_i^{}) - g^{-1}(\boldsymbol { \psi^* }^{\top} A_i X_i^{}+f(X_i))\right]$ equals or is close to $\boldsymbol{0}$, but $\boldsymbol { {\psi}^{*}}$ will in general be different from the true $\boldsymbol { {\psi}}$.

Let ${g^{-1}}^{\prime}$ denote the derivative of the inverse link function $g$. Then  the expectation of $\sum_i^n (U_{1i} - U_{2i})$ conditional on $(X_1, ...X_n)$,  that is,  $\sum_i^n (1- \pi(X_i))X_i w(0, X_i)\left[g^{-1}(f(X_i)) - g^{-1}(  \boldsymbol { \beta }^{\top} X_i)\right]$,  can be written using a Taylor series expansion (function $g^{-1}(f(X_i))$ at the point $\boldsymbol { \beta }^{\top} X_i$) as 
\begin{equation}\label{eq:A.9}
\sum_i^n (1- \pi(X_i))X_i w(0, X_i)\left[{g^{-1}}^{\prime}(\boldsymbol {\beta}^{\top} X_i)(f(X_i) -  \boldsymbol { \beta}^{\top} X_i)+\mathcal{O}[(f(X_i) - \boldsymbol {\beta}^{\top} X_i)^2]\right],
\end{equation}
where the big $\mathcal{O}$ describes the error term in an approximation to the $g^{-1}$ function. The notation $\mathcal{O}[(f(X_i) - \boldsymbol {\beta}^{\top} X_i)^2]$ means the absolute-value of the error of $g^{-1}(f(X_i)) - g^{-1}(  \boldsymbol { \beta }^{\top} X_i) - {g^{-1}}^{\prime}(\boldsymbol {\beta}^{\top} X_i)(f(X_i) -  \boldsymbol { \beta}^{\top} X_i)$ is at most some constant times $(f(X_i) - \boldsymbol {\beta}^{\top} X_i)^2$ when $f(X_i) - \boldsymbol {\beta}^{\top} X_i$ is close enough to 0. Further, the expectation of $\sum_i^n U_{2i}$ conditional on $(X_1, ...X_n)$ is $$\sum_i^n \pi(X_i) X_i w(1, X_i)\left[g^{-1}(\boldsymbol{\psi}^{\top} X_i + f(X_i)) - g^{-1}(\boldsymbol{\psi}^{\top} X_i +  \boldsymbol { \beta }^{\top} X_i)\right],$$ which can be written as 
\begin{equation}\label{eq:A.10}
\sum_i^n \pi(X_i)X_i w(1, X_i)\left[{g^{-1}}^{\prime}(\boldsymbol{\beta}^{\top} X_i +\boldsymbol{\psi}^{\top} X_i)(f(X_i) -  \boldsymbol { \beta}^{\top} X_i)+\mathcal{O}[(f(X_i) - \boldsymbol {\beta}^{\top} X_i)^2]\right].
\end{equation}

Define $\kappa^{*}(A, X) = {g^{-1}}^{\prime}(\boldsymbol{{\beta}^{*} }^{\top} X + \boldsymbol{ \psi^*}^{\top} AX^{})$, where $\boldsymbol{ \psi^*}$ is an assumed limiting value for $\hat{\boldsymbol{ \psi}}$.  Then if weights are defined to satisfy a new balancing criterion $(1- \pi(X)) w(0, X)\kappa^{*}(0, X) = \pi(X) w(1, X)\kappa^{*}(1, X)$, and if the distribution of $X$ is such that the inverse link function is close to linear for the range of $f(X) - \boldsymbol{{\beta}^{*}}^{\top}X$ (so that the Taylor expansion error term is small),  the fact that the expectation of \ref{eq:A.9} is $\boldsymbol{0}$ for $\boldsymbol{\beta} = \boldsymbol{{\beta}^{*}}$ means that the expectation of \ref{eq:A.10} is close to $\boldsymbol{0}$ for $\boldsymbol{\beta} = \boldsymbol{{\beta}^{*}}$. This argument is what was needed to establish the approximate consistency of  the corresponding  new estimator of $\boldsymbol{\psi}$.

Therefore, in single-stage decision settings,  the algorithm for estimation of $\boldsymbol{\psi}$ is concluded as follows:\\
\textbf{Step 1:} Conduct a weighted GLM (e.g., logistic regression) to obtain $\boldsymbol{\hat{\beta}}$ and $\boldsymbol{\hat{\psi}}$. Here the weights are from standard dWOLS weights, such as  $w(a; x) = |a - \mathop{\mathbb{E}[A | X=x]}|$. \\
\textbf{Step 2:} Compute the new weights that satisfy $(1- \pi(X)) w(0, X)\kappa(0, X) = \pi(X) w(1, X)\kappa(1, X)$, where $\kappa(A, X) = {g^{-1}}^{\prime}(\boldsymbol{\hat{\beta}}^{\top} X + \boldsymbol{\hat{\psi}}^{\top} AX^{})$, and ${g^{-1}}^{\prime}$ is identified based on the link function in Step 1. 
For example, the weights can be $$w^{new}(a; x) = |a - \mathop{\mathbb{E}[A | X=x]}| * \kappa(1 - A , X).$$
\textbf{Step 3:} Use the new weights from Step 2, and conduct weighted GLM again, to get a new estimator $\boldsymbol{\widetilde{\beta}}$, and using this, an approximately consistent estimator $\boldsymbol{\widetilde{\psi}}$ of $\boldsymbol{\psi}$.

Remark: In the proof of Theorem 1, we show that the consistency of $\boldsymbol{\widetilde{\psi}}$ depends on the small error term of the first order Taylor expansion of $g^{-1}$ about $\boldsymbol{\beta}^{\top}X$, evaluated at $f(X)$.  This error term will be small when $\boldsymbol{\beta}^{\top}X$ tends to vary in the range where $g^{-1}$ is approximately linear.  In practice, as mentioned in the Methodology section, it may be possible to choose the range of $X$ so that $\boldsymbol{\beta}^{\top}X$ varies in such range, and our estimation will be more precise.  Moreover, the constraint of covariates may have an advantage in terms of the overlap (positivity) assumption for the treatment distributions, and thus the possibility of designing the study to constrain the distribution of $\boldsymbol{\beta}^{\top}X$ is worth exploring more. 



\section{Proof of the uniqueness of $\boldsymbol { {\beta}^{*}}$}\label{Appx.P2B1}
 In Appendix \ref{Appx.P2A1}, in the proof of Theorem 1, we rely on the uniqueness of $\boldsymbol { {\beta}^{*}}$ and large sample theory in \cite{white1982maximum}, and conclude that $\hat{\boldsymbol { \beta }}$ tends to the unique $\boldsymbol { {\beta}^{*}}$ as $n \rightarrow \infty$. Now we are going to prove the uniqueness of $\boldsymbol { {\beta}^{*}}$.

 Note that $\boldsymbol{\beta^{*}}$ is defined as $$\mathbb{E}\left[(1 - \pi(X))Xw^d(0, X)\left[g^{-1}(f(X)) - g^{-1}(  {\boldsymbol { {\beta}^{*}}}^{\top} X)\right] \right]=\boldsymbol{0},$$ and is the root of an analogue of a (vector) score function: $$ \boldsymbol{S} (X; \boldsymbol{\beta}^{}) = \mathbb{E}\left[(1 - \pi(X))Xw^d(0, X)\left[g^{-1}(f(X)) - g^{-1}(  {\boldsymbol{\beta}^{}}^{\top} X)\right] \right].$$ That is, $ \boldsymbol{S} (X; \boldsymbol { {\beta}^{*}}) = \boldsymbol{0}$ with the dWOLS weights that satisfy $(1 - \pi(X))w^d(0, X) =  \pi(X) w^d(1, X)$. For the continuous outcome where the link function is $g(\mu) = \mu$, the parameter $\boldsymbol{\beta}^*$ from $ \boldsymbol{S} (X; \boldsymbol { {\beta}^{*}}) = \boldsymbol{0}$ can be solved explicitly. In the general case, assuming that differentiation with respect to $\boldsymbol{\beta}$ can be carried through the expectation, then the analogue of the Hessian matrix is $$\boldsymbol{H} (\boldsymbol{\beta}) = \pd{\boldsymbol{S} (X; \boldsymbol{\beta}) }{\boldsymbol{\beta}} = - \mathbb{E}\left[\pd{ \left \{(1 - \pi(X))Xw^d(0, X)\left[ g^{-1}(  {\boldsymbol{\beta}^{}}^{\top} X)\right] \right \} }{\boldsymbol{\beta}}\right] =- \mathbb{E}\left[R(0, X)X \kappa(0, X) X^{\top}\right],$$ where $\kappa(A,X) = {g^{-1}}^{\prime}(\boldsymbol {{\beta}^{} }^{\top} X + \boldsymbol{\psi}^{\top}AX)$ and $R(0, X) = (1 - \pi(X))w^d(0, X)$. Then we have the following key corollary to show the uniqueness of $\boldsymbol { {\beta}^{*}}$.
 
 The matrix $\boldsymbol{H} (\boldsymbol{\beta})$ is negative semi-definite for any $\boldsymbol{\beta} \in \mathbb{R}^{p_{}}$ if the link function is monotone increasing, that is, $\kappa(0, X) = {g^{-1}}^{\prime}(\boldsymbol {{\beta}^{} }^{\top} X) > 0$ for any $\boldsymbol {{\beta}^{} }^{\top} X$.
 
 Proof: for any $\boldsymbol{u} \in \mathbb{R}^p$, we have 
 \begin{equation*}
     \boldsymbol{u}^{\top} \boldsymbol{H}(\boldsymbol{\beta}) \boldsymbol{u} = -\mathbb{E}[\sum_{r}^{p}(x_r u_r)^2 R(0, X) \kappa(0, X)],
 \end{equation*}
where $R(0, X)$ is positive. Further, if the first derivative of $g^{-1}$ is always positive, then the above expression $\boldsymbol{u}^{\top} \boldsymbol{H}(\boldsymbol{\beta}) \boldsymbol{u} \leq \boldsymbol{0}$ for all $\boldsymbol{u} \in \mathbb{R}^p$ and   $\boldsymbol{\beta} \in \mathbb{R}^p$ with equality holding when $\boldsymbol{u} = 0.$ Thus, the  matrix $\boldsymbol{H} (\boldsymbol{\beta})$ is negative definite for all $\boldsymbol{\beta}$ and the corresponding log-likelihood function analogue is strictly concave. 
 
 Therefore $\boldsymbol{\beta}^*$ satisfying $ \boldsymbol{S} (X; \boldsymbol { {\beta}^{*}}) = \boldsymbol{0}$  would be the unique root. Note that for the linear treatment-free case, i.e., $f(X) = \boldsymbol{\beta}_{0}^{\top}X$, then we have $\boldsymbol { {\beta}^{*}} = \boldsymbol{\beta}_{0}$.

Note that our dWGLM approach involves two-step regression to consistently estimate the parameter of interest. Each step uses GLM for binary outcomes (e.g., logistic regression) to estimate the parameter. The above argument can be used to show that the first step (logistic) regression provides unique first stage estimates $\boldsymbol{\hat{\beta}}$ and $\boldsymbol{\hat{\psi}}$ for the parameters $\boldsymbol{\beta}$ and $\boldsymbol{\psi}$. Building on these first stage estimates and weights equation (\ref{neww}), we can get new weights, which satisfy weights criterion (\ref{wtcri}) and are always positive. Using the same argument as for the uniqueness of  $\boldsymbol{\beta^*}$, we can prove the uniqueness of the limiting value $\widetilde{\boldsymbol{\beta}^*}$  of the second step parameter denoted as $\widetilde{\boldsymbol{\beta}^*}$ 
when the new weights $w^{new}$ are used. Note that, $\widetilde{\boldsymbol{\beta}^*}$ 
is defined through $\mathbb{E}\left((1 - \pi(X))Xw^{new}(0, X)\left[g^{-1}(f(X)) - g^{-1}(  \widetilde{\boldsymbol { \beta^* }}^{\top} X)\right] \right) = \boldsymbol{0}$.

\section{Derivation of the First-stage True DTR Parameters}\label{Appx.P2C1}
In this section, we derive the true values of the first-stage decision rule parameters in terms of the data generating parameters (\cite{moodie2014q}). Following the notations in Section 3.2 (Study2b), let $f_2 = \theta_0 + \theta_1X_1 + \theta_2O_1 +\theta_3A_1 +\theta_4O_1A_1 +\theta_5X_2 + \varphi_1(X_1) + \varphi_2(X_2)$, then we have
\begin{align*}
    \mathbb{P}( \widetilde{\mathcal{Y}_1} = 1) & = expit \left(f_2 +\left|\theta_{6}+\theta_{7} O_{2}+\theta_{8} A_{1}\right|^{+}\right) \\
&= expit \left(f_2 + O_2A_1 |\phi_1|^{+} + O_2(1-A_1) |\phi_2|^{+} + (1-O_2)A_1 |\phi_3|^{+} + (1-O_2)(1-A_1) |\phi_4|^{+}\right),
\end{align*}  where $|x|^{+} = x * \mathbb{I}(x>0)$, and $\phi_1 = \theta_{6}+\theta_{7}+ \theta_{8}$, $\phi_2 = \theta_{6}+\theta_{7} $, $\phi_3 = \theta_{6}+\theta_{8} $, $\phi_4 = \theta_{6} $ for binary variables  $O_2$ and $A_1$ in $\{0,1\}$. Furthermore,
$\mathbb{E}(O_2 \mid O_1, A_1) = expit(\delta_1O_1 + \delta_2A_1) = 1 - \mathbb{E}(1 - O_2 \mid O_1, A_1),$ thus,
\begin{align*}
    &Q_1(H_1,A_1) =\mathbb{E} (\widetilde{\mathcal{Y}_1} \mid H_1,A_1) \\
    &=  expit \left(f_2 + \mathbb{E}(O_2 \mid H_1, A_1) [A_1 |\phi_1|^{+} + (1-A_1) |\phi_2|^{+}] + \mathbb{E}(1 - O_2 \mid O_1, A_1)[A_1 |\phi_3|^{+} + (1-A_1) |\phi_4|^{+}]\right) 
\end{align*} 


Moreover, 
\begin{align*}
    expit(\delta_1O_1 + \delta_2A_1) &= O_1A_1expit(\delta_1 + \delta_2) + O_1(1-A_1)expit(\delta_1) \\ & {\color{white}(1 - 1)}+ (1-O_1)A_1expit(\delta_2) + (1-O_1)(1-A_1)expit(0) \\ &:= O_1A_1k_1 + O_1(1-A_1)k_2 + (1-O_1)A_1k_3 + (1-O_1)(1-A_1)k_4,
\end{align*}
where $k_1 = expit(\delta_1 + \delta_2), k_2 = expit(\delta_1), k_3 = expit(\delta_2), k_4 = expit(0),$ and $A_1^2 = A_1, (1-A_1)^2 = 1 - A_1, A_1(1 - A_1) = 0.$ Therefore, we have 
\begin{align*}
logit[\mathbb{E} (\widetilde{\mathcal{Y}_1} \mid h_1,a_1)] 
=f_2 &+ |\phi_4|^{+}  + \left(|\phi_3|^{+} - |\phi_4|^{+}\right)A_1  \\&+ O_1A_1k_1\left(|\phi_1|^{+} - |\phi_3|^{+} \right) +  O_1(1-A_1)k_2 \left(|\phi_2|^{+} - |\phi_4|^{+}\right) \\ & + (1-O_1)A_1k_3\left(|\phi_1|^{+} - |\phi_3|^{+} \right)  + (1-O_1)(1-A_1)k_4\left(|\phi_2|^{+} - |\phi_4|^{+} \right).
\end{align*}

Therefore, for the true blip parameters $\boldsymbol{\psi}_1 = (\psi_{10}, \psi_{11})^{\top}$, the above equation gives the coefficient of $A_1$ as $$\psi_{10} = \theta_3 + |\phi_3|^{+} - |\phi_4|^{+} + k_3 \left(|\phi_1|^{+} - |\phi_3|^{+} \right) - k_4 \left(|\phi_2|^{+} - |\phi_4|^{+} \right),$$ and the coefficient of $O_1A_1$ as $$\psi_{11} = \theta_4 + (k_1-k_3) \left(|\phi_1|^{+} - |\phi_3|^{+} \right) - (k_2 - k_4) \left(|\phi_2|^{+} - |\phi_4|^{+} \right).$$

\end{appendix}

\end{document}